\documentclass[aps,pop,nobibnotes,preprint,superscriptaddress,showkeys,amsmath,floatfix]{revtex4-1}

\usepackage[colorlinks]{hyperref}

\usepackage{siunitx}

\usepackage{graphicx}
\usepackage{import}
\usepackage{xcolor}
\usepackage{textcomp}
\usepackage[capitalize]{cleveref}
\usepackage{placeins}
\usepackage[utf8]{inputenc}
\usepackage{xstring}
\usepackage{letltxmacro}
\usepackage{ulem}

\usepackage{soul} 
\soulregister\cite7
\soulregister\ref7
\soulregister\pageref7
\soulregister\cref7
\soulregister\Cref7

\usepackage{amssymb,amsfonts,amstext,graphics,graphicx,subfigure}
\usepackage{color}

\newcommand{\beginsupplement}{%
        \setcounter{table}{0}
        \renewcommand{\thetable}{S\arabic{table}}%
        \setcounter{figure}{0}
        \renewcommand{\thefigure}{S\arabic{figure}}%
     }
\date{\today}

\newcommand\hcancel[2][0.5pt]{%
  \ifmmode\sbox\CBox{$#2$}\else\sbox\CBox{#2}\fi%
  \makebox[0pt][l]{\usebox\CBox}%
  \rule[0.5\ht\CBox-#1/2]{\wd\CBox}{#1}}

\begin{document}

\title{Elastoinertial effects govern dynamic response of soft hair beds}
\author{Jonas Smucker}
\author{Naiesa Freeman}
\author{Eric Caballero}
\affiliation{Department of Physics, University of Texas at Austin}

\author{Philip J. Morrison}
\affiliation{Department of Physics, University of Texas at Austin}
\affiliation{Institute for Fusion Studies, University of Texas at Austin}

\author{Jos\'e Alvarado}
\email{alv@chaos.utexas.edu}
\affiliation{Department of Physics, University of Texas at Austin}
\maketitle
\section{Abstract}

Fluid-immersed hair beds are ubiquitous in biology — from the endothelial glycocalyx and primary cilia to intestinal microvilli — where they serve as mechanosensors that transduce dynamic flow signals into biochemical regulatory responses.
Despite the inherently dynamic nature of physiological flows, the dynamic mechanical properties of fluid-immersed hair beds under time-varying conditions remain poorly characterized.
Here we investigate the transient rheological response of elastic hair beds to large-amplitude oscillatory shear flows at low to intermediate Reynolds number.
While the hairs and fluid themselves obey linear constitutive laws, their coupled interaction produces a dynamic nonlinear response that depends sensitively on driving frequency and amplitude.
We identify a crossover from a stress-lagging regime to a stress-leading regime, which is governed by an interplay between fluid viscosity, fluid inertia, and hair elasticity.
A simplified rigid-beam model qualitatively captures the crossover behavior.
Characterizing the dynamic flow response of soft hair beds has direct biological implications, since the lag time  sensitively determines the stability of mechanosensory signaling in the feedback loops underlying essential biological processes such as vasodilation, ciliary remodeling, and tubular reabsorption.
Our results establish a framework for understanding how the physical properties of biological hair beds optimize dynamic information transmission during mechanotransduction.

\section{Introduction}

Fluid-immersed hairs and hair beds transduce dynamic mechanical information across a remarkably wide range of biological systems.
Whiskers, stereocilia, filiform hairs, trichobothria, primary cilia, and the endothelial glycocalyx all deform in response to transient fluid forcing, triggering downstream biochemical signals that govern behavior and adaptation \cite{Fettiplace-2014-Physiol.Rev., Boublil-2021-Sensors}.
Although these systems share common, hair-like geometries, they serve a diverse set of functions, transducing mechanical information from upstream time-varying flows to downstream regulatory processes.
Several well-characterized biological examples illustrate this functional role.
Epithelial cells lining kidney tubules and bile ducts extend rod-like primary cilia, microtubule-based projections of lengths \qtyrange{1}{10}{\micro\meter}, that deflect under luminal flow and initiate intracellular calcium and nitric oxide (NO) signaling cascades \cite{satir_overview_2007, spasic_primary_2017, saternos_primary_2020}.
The resulting NO release drives vasodilation and regulates blood pressure \cite{kadekaro_centrally_2000, rees_role_1989, boo_flow-dependent_2003}.
Blood vessel walls are coated with the endothelial glycocalyx, a passive brush-like structure containing heparan sulfate-tipped core protein filaments arranged in a near-periodic array, with lengths of approx. \qty{100}{\nano\meter} \cite{weinbaum_mechanotransduction_2003}.
Shear-induced bending of these filaments similarly drives NO production \cite{bartosch_endothelial_2017}.
In fish and aquatic amphibians, epithelial neuromast cells of the lateral line system couple mechanosensory hair cells to a gelatinous cupula that deflects in response to water flow, with cupular geometry and stiffness setting the frequency range of detection \cite{vanNetten-1987-HearingResearch, McHenry-2008-JCompPhysiolA, vanNetten-2014-TheLateralLineSystem}.
Therefore, the mechanical properties of mechanosensing hair beds are critical to their function.

But organisms are rarely stationary, and the flows that impart stresses on hair beds are rarely steady.
Signals from deformed hair beds drive feedback loops that regulate behavior and adaptation across multiple timescales.
On short timescales of seconds, glycocalyx and ciliary deformation drive NO production and vasodilation \cite{weinbaum_mechanotransduction_2003, bartosch_endothelial_2017, spasic_primary_2017}.
Meanwhile on longer timescales of minutes to hours, primary cilia actively remodel in response to cumulative flow stimulation. On these timescales, cilia length can vary by a factor of two through an ion-concentration-dependent transport process \cite{besschetnova_identification_2010}.
Mechanical stimulation also modifies ciliary stiffness through a negative feedback loop \cite{nguyen_primary_2015}.
Shorter, stiffer cilia deflect less — a response that may protect against cleavage under large stresses \cite{spasic_primary_2017}.
Furthermore, misregulation of mechanotransduction pathways underlies disease.
Loss of ciliary flow sensing is implicated in autosomal dominant polycystic kidney disease via the polycystin-1/2 pathway \cite{Nauli-2003-NatGenet}.
Impaired glycocalyx-mediated NO production is an early indicator of vascular pathology \cite{boo_flow-dependent_2003}.
In addition, the timing of mechanosensory signals is consequential.
A well-known fact of feedback loops is that the lag between a stimulus and a downstream response largely determines stability.
Time delays are well-established as destabilizing in control systems arising in robotics and physiology alike \cite{milton_time-delayed_2009, landry_dynamics_2005, martin_predator-prey_2001, bechhoefer_control_2021}.
The coupled dynamics between fluid flows and hair-bed deformations therefore directly govern the fidelity and stability of mechanosensory regulation, and are an important area of study.

Previous studies have characterized the static mechanical response of hair beds to steady flows.
The deformation of single fibers under viscous flow has been studied experimentally \cite{wexler_bending_2013, Duprat-2014-LabChip}, while subsequent experiments on hair beds revealed nonlinear drag reduction arising from elastoviscous coupling \cite{alvarado_nonlinear_2017}.
Building on prior models \cite{gopinath_elastohydrodynamics_2011, young_dynamics_2012}, analytical and computational models have further characterized flow responses \cite{stein_coarse_2019, smucker_integrability_2022, rahimi_2022, Pang-2025-PhysicsofFluids, Sun-2025-PhysicsofFluids}.
Additional experiments have explored collective stiffening \cite{thomazo_collective_2020}, lubrication \cite{thomazoProbingInmouthTexture2019, Peng-2021-ExpMecha}, and hair-hair contact during drainage \cite{ushay_interfacial_2023}. More recent studies have extended these studies to Poiseuille flows \cite{Jambon-Puillet-2026-J.FluidMech., Jammalamadaka-2026-} and to transient responses to pulsed flows in the glycocalyx \cite{mitsoulas_2022}.
The mechanical properties of neuromast cupulae have been characterized and shown to sensitively determine deformation and hence mechanotransduction \cite{McHenry-2007-JExpBiol, McHenry-2008-JCompPhysiolA}.
Despite this progress, the nonlinear response of hair beds to time-varying, periodic flows— the regime most relevant to biological flows— remains largely unexplored.

Here we address this gap by experimentally characterizing the response of a biomimetic hair bed to large-amplitude oscillatory shear flows.
We quantify the dynamic response using the mechanical impedance, a complex quantity capturing both the amplitude and phase of the stress-velocity relationship.
We identify an interplay between fluid inertia, fluid viscosity, and hair elasticity that governs whether the transmitted stress lags or leads the imposing oscillatory flow.
A simplified model treating the hairs as rigid beams coupled with a nonlinear torsional spring recapitulates the key features of the response.
Understanding this dynamic behavior is an essential component in connecting the fundamental fluid-structure interactions of fluid-immersed hair beds to the fidelity of mechanotransduction, the stability of associated feedback loops, and ultimately the behavior and adaptation of living systems.

\section{Results}

\subsection{Hair bending decreases drag}

\begin{figure}
\centering
    \includegraphics[scale=.5]{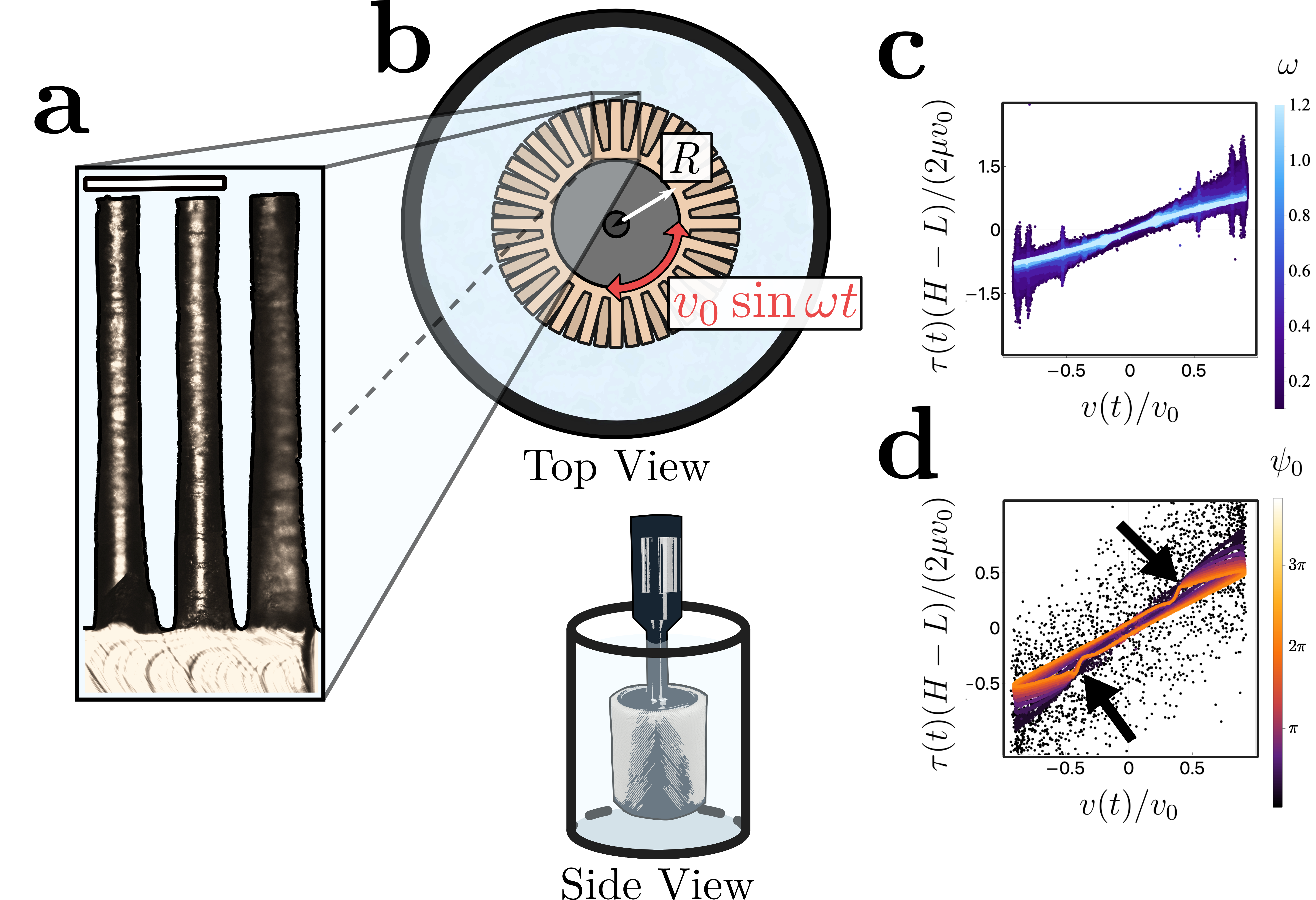}
    \caption{\textbf{Experimental setup and stress-velocity Lissajous diagrams}. (\textbf{a}) False-color microscope image of three hairs. Scale bar 1 mm. (\textbf{b}) Top and side views of the inner cylinder of the Taylor-couette geometry used in rheology experiments. (\textbf{c}) Scaled stress-velocity output is shown for a frequency sweep at fixed $\psi_0=10$. (\textbf{d}) Scaled amplitude sweep at fixed $\omega=12\,\mathrm{Hz}$. The experimental data shown in panels (c) and (d) are taken from a hair bed with $L=3.67\,\mathrm{mm}$, $\phi=0.043$, and $\mu=1\,\mathrm{Pa\,s}.$
    }
    \label{fig:expSetup}
\end{figure}

In order to determine the response of soft hair beds to time-varying, external shear flows, we conduct oscillatory rheology experiments. In short, we mount soft hair beds onto a Taylor-Couette rotor and submerge in silicone oil (see Methods and \cref{fig:expSetup}a,b). We specify the velocity of the rotor, parametrized by the velocity $v(t) = v_0 e^{\mathrm{i}\omega t} = \omega\psi_0(R+L) e^{\mathrm{i}\omega t}$ of undeformed hair tips, with $R$ the rotor radius and $L$ the hair length. We systematically vary the frequency $\omega$ (\cref{fig:expSetup}c) and amplitude $\psi_0$ (\cref{fig:expSetup}d) of rotor oscillation and record the resulting stress $\tau(v)$. If the system were purely Newtonian, we would expect the Lissajous figures to be completely linear; indeed, we observe a mostly linear dependence between $\tau$ and $v_0$ (\cref{fig:expSetup}c,d), along with measurement noise at low frequencies (panel c, dark curves) and low amplitudes (panel d, dark curves). However, we also observe deviations from linear behavior, including kinks (panel d; arrows) and non-zero enclosed areas (panel d; bright curves). These observations suggest that hair deformation contributes to the rheological response at high frequencies and amplitudes of shear deformation. Furthermore, the Reynolds number $\mathrm{Re}\equiv\frac{\rho v_0 H}{\mu}$ of the driving shear flows lies in the range $(0.0001, 20)$, well within the laminar regime but high enough where inertial effects may play a role; $\rho$ is the density and $\mu$ is the viscosity of the surrounding fluid.

In order to further characterize the rheological response of hair beds, we consider the (dynamic, area-specific) impedance $Z \equiv \tau(t)/v(t)$, with shear stress $\tau = \frac{\mu v(t)}{H - h(t)}$ and $h$ the height of the hair tip from the base. In the weakly deformed case, $Z_0 \equiv Z(v\rightarrow 0) = \mu/(H-L)$. As the rotor velocity increases, hairs bend down, $h(t)$ decreases toward zero, and $Z$ approaches the limiting value $Z_\infty \equiv Z(v \rightarrow \infty) = \frac{\mu}{H}$. As a result, $Z$ can be viewed as a fluidic ``resistance'' to Couette flow that depends sensitively on geometric nonlinearity. In this article, we will parametrize the complex-valued impedance by its magnitude $|Z|$ and argument $\arg(Z)$. The magnitude directly corresponds to the total fluidic ``resistance'' to Couette flow, and the argument corresponds to the phase lag between driving boundary velocity and resulting shear stress measured at the base of the hairs. The impedance $Z$ is related to the shear modulus $G = \tau/\gamma = \mathrm{i} \omega R / k_\gamma Z$, where $\gamma = k_\gamma \psi$ is the fluid strain and $k_\gamma$ is a constant that depends on the geometry (Methods). 

\begin{figure}
    \centering
    \includegraphics[scale=.8]{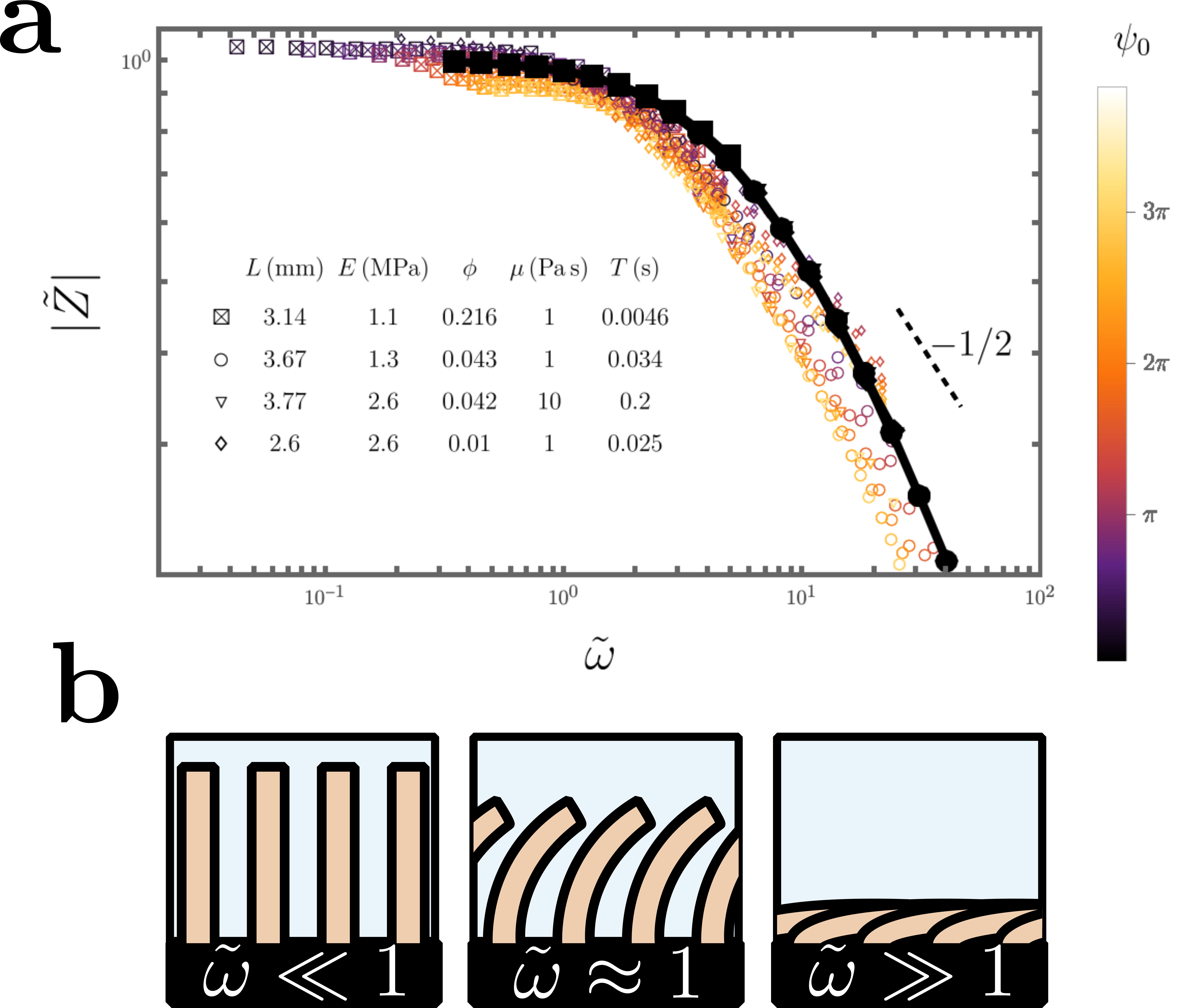}
\caption{\textbf{The dynamic response of hairs demonstrates elastoviscous drag reduction}. (\textbf{a}) Experimentally measured data (purple to yellow symbols) and numerical results from the theoretical model (black symbols and line) of the magnitude of the rescaled impedance $\tilde{Z}=\frac{Z-Z_\infty}{Z_0-Z_\infty}$ versus the rescaled frequency $\tilde{\omega}\equiv\frac{4\mu L^2 v_0}{E\phi a^2 H}\left(1-\frac{L}{H}\right)^{-3/2}$, with $v_0=(R+L) \psi_0 \omega$. Symbol shape denotes hair bed and fluid properties (table). (\textbf{b}) Schematic of the three observed regimes: weakly deformed ($\tilde Z \sim \tilde \omega^0$, left), reconfiguration ($\tilde{\omega}\approx 1$, center), and strongly deformed ($\tilde Z \sim \tilde \omega^{-1/2}$, right).}
   \label{fig:rezs}
\end{figure}

Shear flows deform hairs and reduce the magnitude of the impedance, resulting in a drag-reducing nonlinearity. This can be seen by transforming stress and velocity data to two dimensionless quantities. First, we consider the rescaled impedance $\tilde{Z} = \frac{Z-Z_\infty}{Z_0-Z_\infty}$. We also consider the rescaled frequency $\tilde{\omega}\equiv\frac{4\mu L^2 v_0}{E\phi a^2 H}\left(1-\frac{L}{H}\right)^{-3/2}$, an elastoviscous parameter that expresses a ratio between the viscous and elastic effects resulting from the imposed oscillatory flow. A perturbation analysis for small deformation reveals that the rescaled impedance depends on $\tilde{\omega}$ \cite{alvarado_nonlinear_2017}. Plotting the magnitude $|\tilde{Z}(\tilde{\omega})|$ (\cref{fig:rezs}a; symbols) demonstrates that our set of experiments, with different sets of parameters and driven with different frequencies and amplitudes, all collapse onto approximately the same functional form. The dataset with the sparsest packing density of hairs ($\phi = 0.01$; \cref{fig:rezs}a, diamonds) appears to deviate more strongly from the other datasets, likely due to shear flows penetrating the hair bed. We observe that weak driving (low values of $\tilde{\omega}$) do not significantly deform hairs, and the impedance is roughly constant($|\tilde{Z}| \approx 1$; $|Z| \approx Z_0)$. As shear flows become stronger ($\tilde{\omega}$ increases), $|\tilde Z|$ begins to decrease past a value of $\tilde{\omega} \approx 1$, where viscous and elastic effects are comparable. This decrease in drag occurs because hair bending increases the gap width $H-h(t)$. As shear flows become very large ($\tilde{\omega} \gg 1$), impedance tends toward zero, and we recover a power law $|\tilde{Z}| \sim \tilde{\omega}^{-1/2}$.

These results are consistent with the steady flow case \cite{alvarado_nonlinear_2017}, where hair beds were subject to constant shear rate. In the strongly deformed limit, hair curvature is concentrated at the base, and otherwise mostly straight \cite{smucker_integrability_2022}. These studies successfully modeled hair-bed response by treating hairs as continuous Kirchoff rods with infinite degrees of freedom, parametrized by a curvilinear coordinate $s$ along the hair's backbone \cite{audoly_elasticity_2010}. We have attempted to develop a similar model for hair beds subject to time-varying flows, but found this approach to be intractable. Instead, in order to simplify the problem, we develop a model which approximates flexible fibers as rigid rods with nonlinear torsional springs.

\subsection{Rigid beam model recovers drag-reducing nonlinearity}
\label{sec:model}

\begin{figure}
    \centering
    \includegraphics[scale=.7]{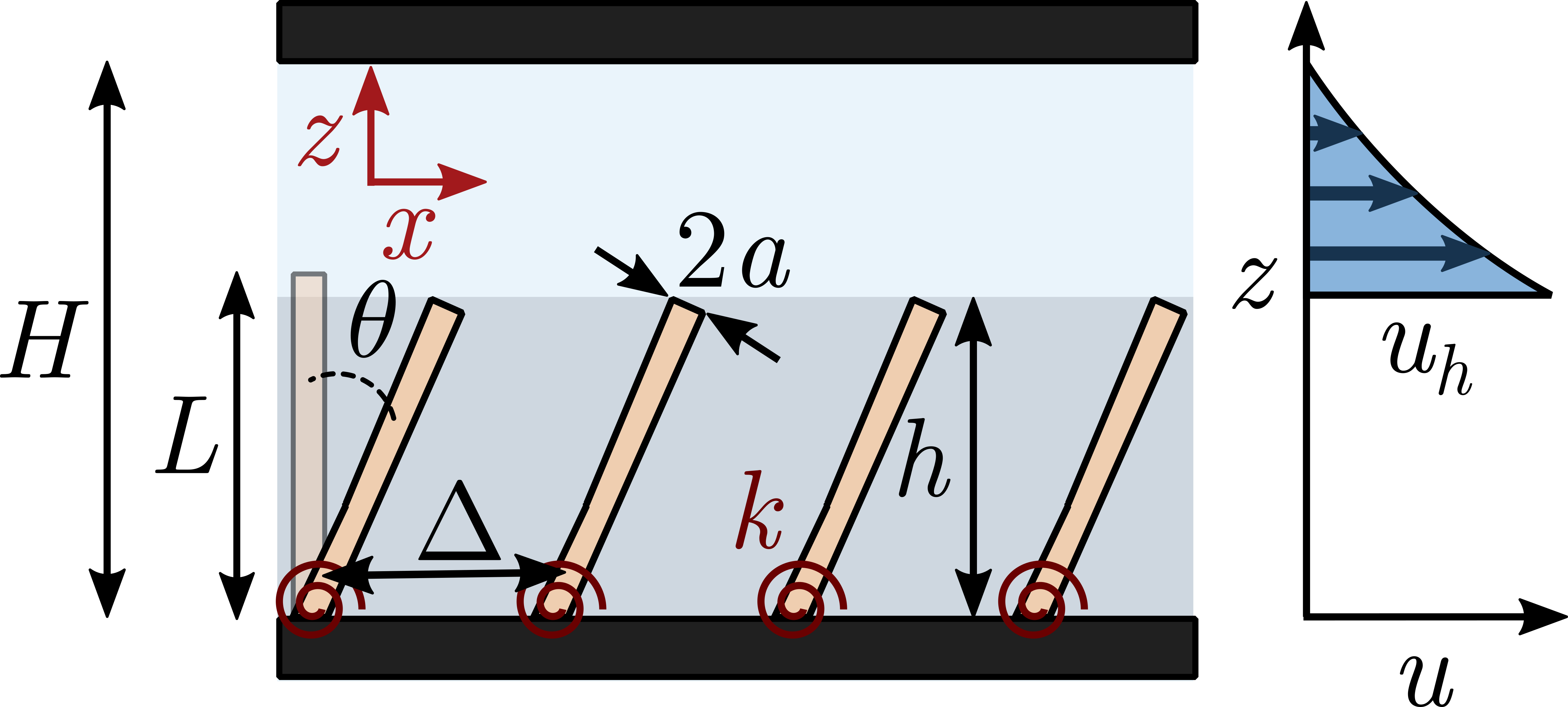}
   \caption{\textbf{Rigid link model}. Flexible hair beds are modeled as beds of rigid rods with pivot anchors at the base and nonlinear torsional springs. Rods have length $L$, diameter $2a$, angle $\theta$ with respect to the vertical, and spacing $\Delta$. The torsional spring has spring constant $k$, which is a function of the velocity $v$ of hairs. The fluid has viscosity $\mu$, velocity $u(z)$, and subject to no-slip boundary conditions at the hair-tip plane, with $u(z=h) = v$.}
   \label{fig:model}
\end{figure}

In order to develop a model for the complex dynamic impedance $\tilde{Z}(\tilde{\omega})$, we model hairs as rigid rods of length $L$ and angle $\theta$ with respect to the surface normal (\cref{fig:model}).
Rods are anchored at the surface with a pivot, as opposed to a cantilever for our earlier flexible fiber models \cite{alvarado_nonlinear_2017, smucker_integrability_2022}.
When a force is applied to the rod, its restoring torque is therefore not provided by a bending moment.
Rather, we approximate the coupled fluid-structure interaction with a torsional spring whose restoring torque $k \sin\theta$ contains two terms: a constant term reflecting linear response and nonlinear term that scales with the square root of the velocity:

$$\frac{k(\hat v)}{Ea^4L^{-1}} = \frac{\pi}{2^{3/2}} \left(\frac{\sqrt{15}}{2} + \hat v^{1/2} \right).$$

Here, $k$ is the spring constant (measured in units of \unit{\newton\meter}), $E$ the rod's elastic modulus, $a$ the cylindrical rod's radius, $L$ its length, $\epsilon = L/H$ the dimensionless hair length, and $\hat v$ the dimensionless velocity of the opposing surface (Supplementary Information).
We further describe a bed of rods with rod-rod spacing $\Delta$ and, equivalently, area packing fraction $\phi=\frac{2\pi}{\sqrt{3}}\frac{a^2}{\Delta^2}$.
We assume that all rods in the bed deform identically, thus we can consider the response of a single rod.
The rotor imposes a fluid velocity $u(h(t))\equiv u_h$ at the hair-tip plane which is located a distance $h(t) \equiv L \cos \theta(t)$ away from the hair-base.
With these assumptions, we solve the unsteady Stokes equations between the rod-tip plane and the moving top plate.
We then compute the torque at the rod tip and balance it with the restoring torque from the rotational spring, yielding:

\begin{equation}
    k(v)\sin\theta(t)=-\frac{\pi a^2}{\phi}\tau_h(t)L\cos\theta(t)
    \label{eq:ode}
\end{equation}
for the hair and 
\begin{align}
u_t(z,t)&=\nu u_{zz}(z,t),\\
u(h(t),t)&=(R+h(t))\omega\psi_0\sin\omega t-\dot{\theta}(t)L\cos\theta(t),\\
u(H,t)&=0
\end{align} 
for the fluid. 
Letting $\xi(t)=\frac{z-h(t)}{H-h(t)}$ and $v(\xi,t)=\frac{u(z(\xi),t)}{(R+L)\psi_0\omega}$ results in the following system of equations
\begin{align}
\label{eq:system}
\begin{split}
\dot{v}&=\frac{\dot{h}}{H-h}(1-\xi)v'+\frac{\nu}{(H-h)^2}v''\\
v(0,t)&=\frac{R+h}{R+L}\sin\omega t-\frac{L}
{\psi_0\omega(R+L)}\dot{\theta}\cos\theta\\
v(1,t)&=0\\
\tan\theta&=-\psi_0\omega T \frac{H(1-\epsilon)}{H-h}v'(0,t)\left( 1 + \frac{2}{\sqrt{15}} |v'(0,t)|^{1/2} \right)^{-1}
\end{split}
\end{align}

where $T\equiv\sqrt{\frac{32}{15}}\left(\frac{L}{a}\right)^2\phi^{-1}(1-\epsilon)^{-1}\left(\frac{R+L}{H}\right)\frac{\mu}{E}$ is a characteristic timescale for the hairs.
This timescale was obtained upon non-dimensionalization of the governing equations and it diverges when the total hair length approaches the channel height ($\epsilon \rightarrow 1$; see Discussion). The primes and dots indicate derivatives with respect to $\xi$ and $t$, respectively.

We numerically solve these equations and compute $|\tilde{Z}(\tilde{\omega})|$ (\cref{fig:rezs}a; black lines and symbols). We find reasonable agreement between model and experiment, confirming that the dynamic rheological response we observe is described by a mutual interplay between elasticity, viscosity, and fluid inertia. Furthermore, we observe that not all predicted curves collapse exactly upon each other, likely because our simplified model contains only one spatial degree of freedom. We furthermore note here that the dataset with high viscosity ($\mu = 10 \ \mathrm{Pa \cdot s}$; \cref{fig:rezs}a, triangles) agrees with the other datasets and with our model, suggesting that secondary viscous flows from neighboring hair deformations do not significantly affect $|\tilde Z|$ (see Discussion).

\subsection{Elastoinertial effects sensitively determine time lag}

\begin{figure}
    \centering
    \includegraphics[scale=.8]{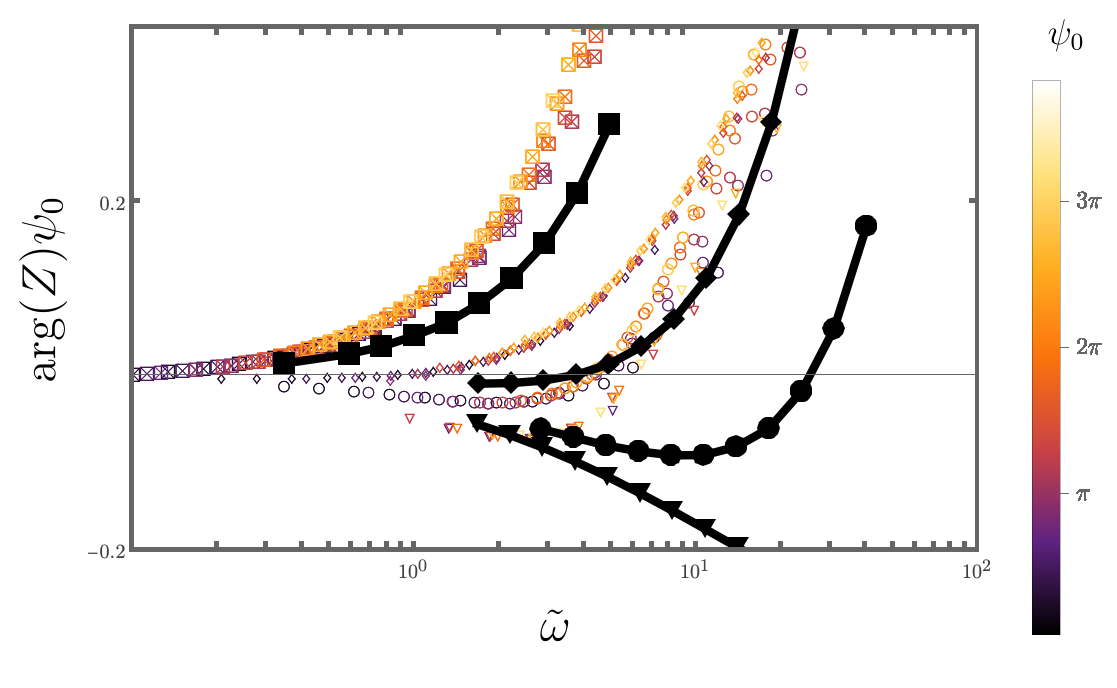}
   \caption{\textbf{Measured stress can either lag or lead the imposed velocity}. Experimentally measured data (purple to yellow symbols) and numerical results from the theoretical model (black symbols and line) of the argument of the impedance $\arg{Z}$ versus $\tilde{\omega}=\frac{4\mu L^2 v_0}{E\phi a^2 H}\left(1-\epsilon\right)^{-3/2}$.}
   \label{fig:imzs}
\end{figure}

\begin{figure}
    \centering
    \includegraphics[scale=.8]{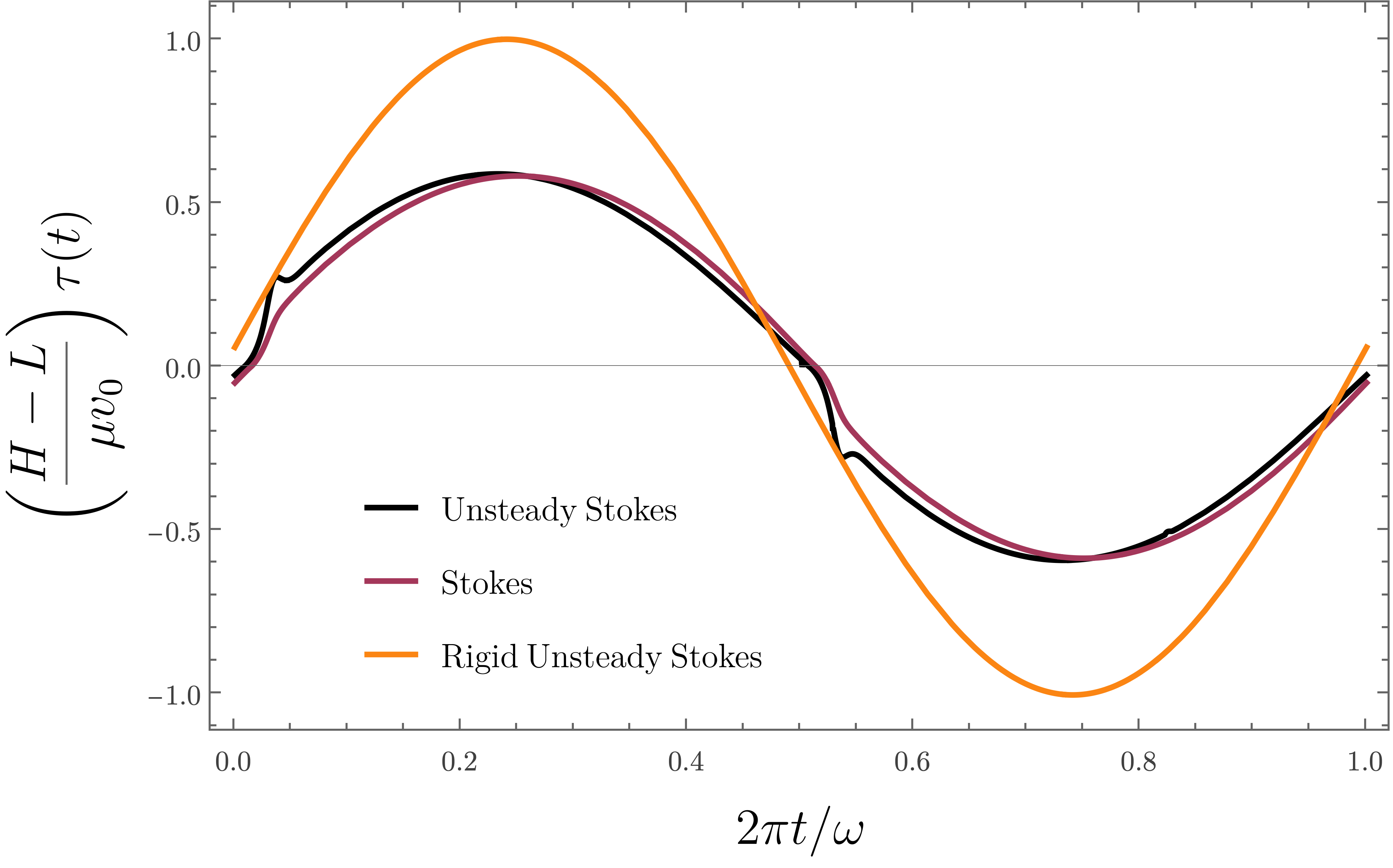}
   \caption{\textbf{Elastoinertial effects underlie nonlinear response}. Time trajectories of shear stress at the base of hairs $\tau(z=0)$, normalized by the characteristic scale $\frac{\mu v_0}{H-L}$ for three models. Orange curve: model with undeformable hairs but including fluid inertial effects. Purple curve: model with finite restoring torque but lacking the unsteady term. Black curve: model with both deformable hairs and fluid inertia.}
   \label{fig:stressTSeries}
\end{figure}

So far we have investigated drag-reduction in hair beds using rheological experiments and our model by considering the magnitude of $\tilde Z$. Drag reduction is also reflected in our measured Lissajous figures (cf. \cref{fig:expSetup}), where the slope of the line traced decreases as frequency or amplitude increase. However, we also observed i) a non-zero enclosed area at high amplitude and ii) small kinks. These observations suggest that $Z$ can have an out-of-phase component (i), corresponding to a time lag between applied boundary velocity and measured shear at the base of the hairs. Brief bouts of nonlinear behavior (ii) may underlie this time lag. Neither of these effects appear to be described by the magnitude $|\tilde{Z}|$. This is reflected in our observation that the real part of $Z$ is at least one order of magnitude larger than the imaginary part (\cref{fig:reZControl,fig:imZControl}).

\begin{figure}
    \centering
    \includegraphics[scale=.7]{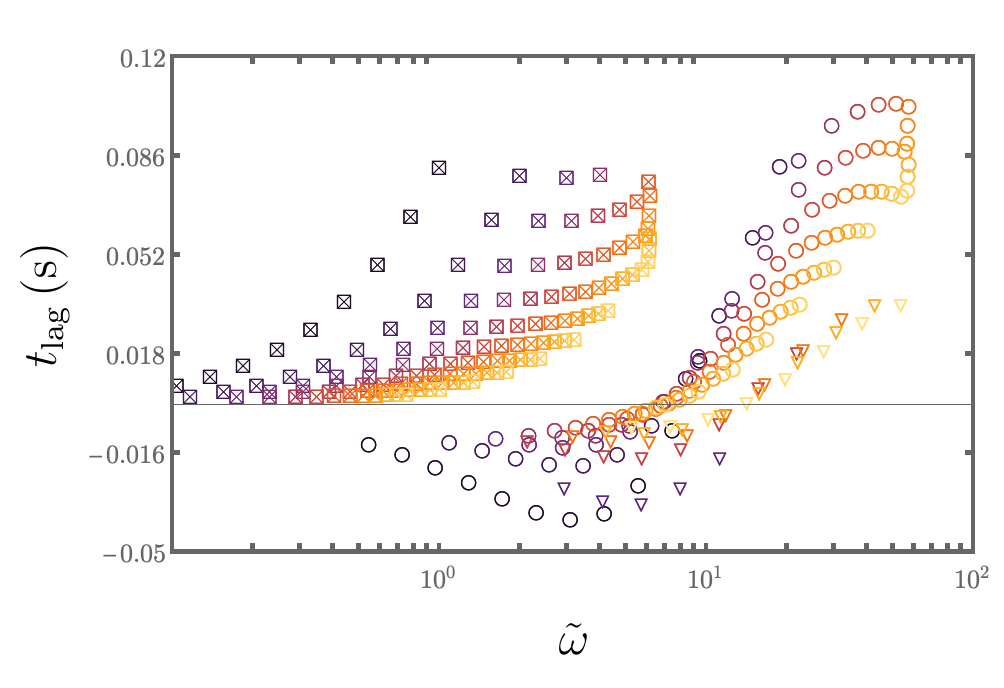}
   \caption{Lag time $t_\mathrm{lag}$ as a function of rescaled frequency $\tilde \omega$ (cf. \cref{fig:rezs})}
   \label{fig:tLag}
\end{figure}

Therefore, in order to further characterize the dynamic response of hair beds, we now consider the argument of the impedance $\arg(Z)$, which is related to the time lag $\tau_{lag}$ by $\arg(Z)=-\omega\tau_{lag}$. Negative $\arg(Z)$ corresponds to a positive lag between stress and velocity while positive $\arg(Z)$ corresponds to negative lag time, or lead. In the Stokes limit (absence of fluid inertia), hair-tip position lags the imposed velocity and the argument is positive. We observe that the magnitude of the time lag between stress and velocity is always less than $0.1$s in our experiments (\cref{fig:tLag}).

We begin by computing $\arg(Z)$ for our experiments (\cref{fig:imzs}, purple-to-yellow symbols).
For all four datasets, in the zero frequency limit, we observe no lag.
This result reflects the quasistatic nature of low-frequency driving.
Yet as frequency increases, we observe that $\arg{Z}$ curves do not collapse onto one master curve, as was the case with $|\tilde Z|$.
(Interestingly, we do find that inserting a prefactor of $\psi_0$ on the ordinate axis assists with collapse within datasets.)
Instead, we observe two distinct functional forms.
Two datasets (\cref{fig:imzs}a, circles and triangles), first clearly decrease to negative values (positive lag) and then increase to positive values above some crossover frequency.
For two other datasets (\cref{fig:imzs}a, squares and diamonds), we find a functional form that appears to increase from values close to zero toward clearly positive values.
We note that the dataset with the highest viscosity (triangles) exhibits the lowest observed values of $\arg{Z}$, consistent with slow relaxation in the limit of overdamped dynamics.
Meanwhile, positive $\arg{Z}$ corresponds to negative lag.
This surprising observation suggests that a nonlinear interplay between hair elasticity and fluid inertia governs phase-lead behavior.

Our model qualitatively agrees with experiment (\cref{fig:imzs}a, black lines and symbols).
It captures the two classes of functional forms we see in experiment.
Agreement is closest at low $\tilde{\omega}$.
However, experiment and model increasingly disagree as the forcing parameter increases.
Furthermore, disagreement is highest for the high-viscosity sample, where $\mu=\qty{10}{\pascal\second}$ (triangles).
Collective stiffening \cite{thomazo_collective_2020} and secondary flow effects may dominate in this regime, which our model neglects.

\subsection{Elastoinertial exchange induces hair ``bounce''}

According to our model, the stresses measured by our rheology experiments are governed by the gap height $H-h$ between the hair-tip plane and the opposing surface.
Recalling the kinks observed in the Lissajous figures (cf. \cref{fig:expSetup}d), we ask whether these kinks relate to elastoinertial effects.
To this end, we plot our model's predicted stress $\tau$ over one cycle under three different assumptions (\cref{fig:stressTSeries}).
First, we consider the case where hairs are undeformable, but fluid inertial effects are considered (orange curve).
We observe an oscillatory shear stress governed largely by a single mode.
Next, a model that considers hair deformation but lacks the unsteady inertial term (purple curve) also appears to be governed by a single mode, although minor deviations appear to occur at times $t \approx 0.05$ and $t \approx 0.55$.
Finally, a model that considers both hair elastic deformation and fluid inertial effects (black curve) demonstrates two pronounced spikes at $t \approx 0.05$ and $t \approx 0.55$.
These results strongly suggest that elastoinertial effects indeed govern a nonlinear response, which manifests as the presence of intermittent perturbations.

\begin{figure}
    \centering
    \includegraphics[scale=.8]{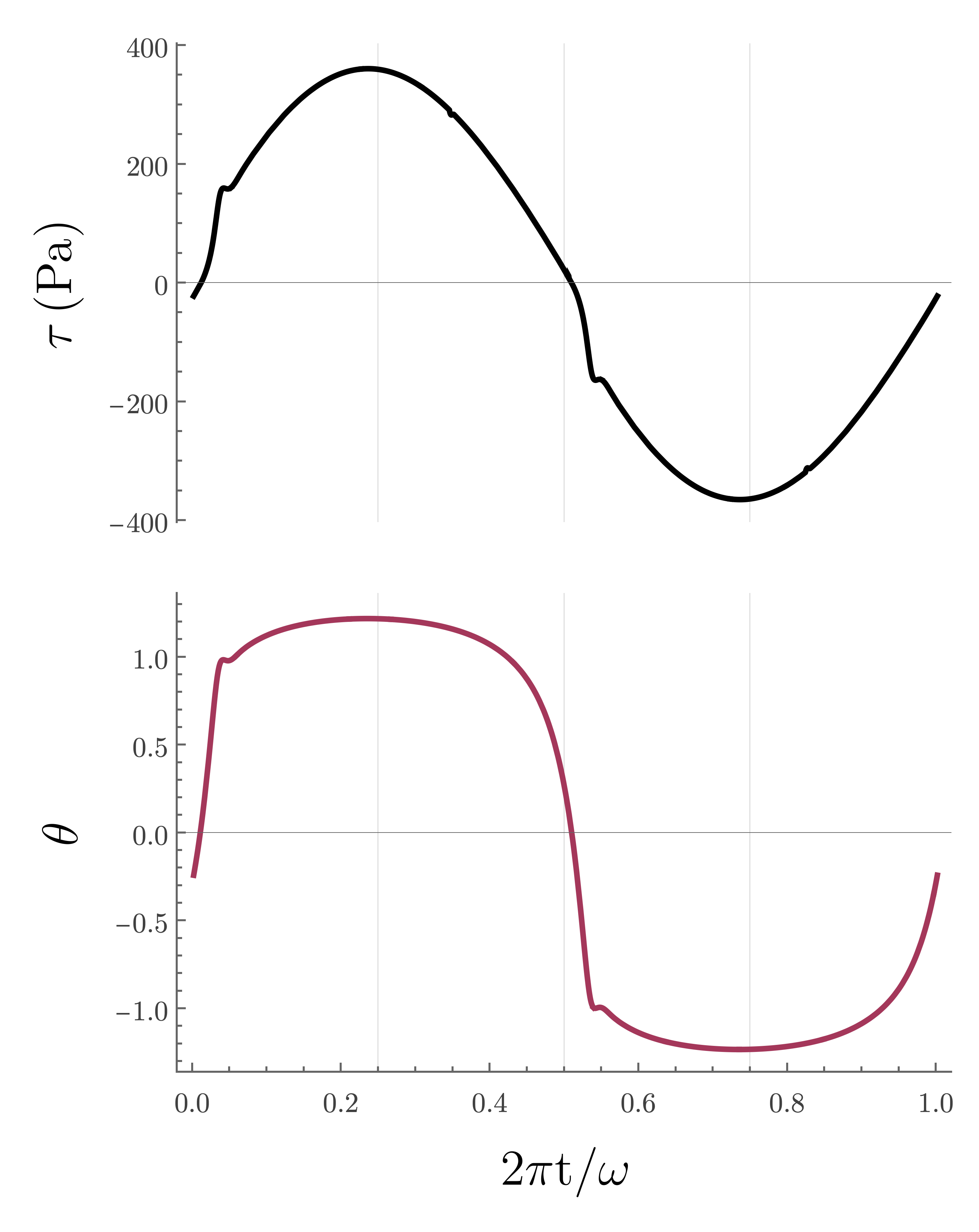}
   \caption{\textbf{Hairs ``bounce" near the minima in their height}. Top: Trajectory of shear stress at the hair tips $\tau(z=h)$ versus time for one cycle. Bottom: Trajectory of hair angle $\theta$ versus time for one cycle.}
   \label{fig:profiles}
\end{figure}

Within the context of our model, the localized perturbations in stress occur completely due to changes in the hair height.
To illustrate this point, we plot the hair angle over the course of one cycle (\cref{fig:profiles}).
We again observe the same localized peaks as with the stress, also located at times  $t \approx 0.05$ and $t \approx 0.55$.
Hairs therefore ``bounce'' due to elastoinertial effects.
Given that hair elasticity and fluid inertia offer two independent energy storage modes, a rich exchange of these two modes most likely underlies the bouncing phenomenon and the kinked Lissajous figures we observe.

\section{Discussion}

We have studied the problem of hair beds subject to time-dependent fluid flows through oscillatory rheology experiments (cf. \cref{fig:expSetup}).
We characterized the phase and magnitude of the stress-velocity response with the complex impedance $Z$ (cf. \cref{fig:rezs,fig:imzs}).
The magnitude of the rescaled impedance $|\tilde Z|$ largely agrees with that of the stationary flow problem (cf. \cref{fig:rezs} and \cite{alvarado_nonlinear_2017}).
The argument $\arg{Z}$ takes on negative values (stress lagging velocity) as well as positive values (stress leading velocity).
Our model found that a nonlinear coupling between elasticity, inertia, and viscosity sensitively determines the phase of the transmitted stress, including its sign.
We now discuss the implications of this finding for biological mechanosensing and feedback control, the limitations of our model, and potential engineering applications.

In the context of living systems, fluid flows impart shear stresses on hair beds, induce bending moments along the deformed hairs, and activate mechanoreceptors at the cell surface.
The timescales with which hairs respond to fluid flows are certain to affect the lag time of associated feedback loops.
In particular, lag between upstream fluid flows and stress at the base can delay feedback, which can destabilize control systems \cite{milton_time-delayed_2009, landry_dynamics_2005, martin_predator-prey_2001, bechhoefer_control_2021}.
Our results demonstrate that lag time depends sensitively on the forcing parameter $\tilde\omega$.
The physical properties and geometry of the fluid and hair bed could directly govern the stability of the feedback loop of critical biological functions.
For example, blood pressure is regulated in the vasculature by controlling vessel diameter through vasodilation and vasoconstriction.
The glycocalyx participates in this loop by transducing wall shear stress into NO production, which drives smooth muscle relaxation \cite{weinbaum_mechanotransduction_2003, bartosch_endothelial_2017, spasic_primary_2017}.
Meanwhile for primary cilia, bending moments transmitted to the cell base can activate ion channels, potentially allowing calcium and other signaling molecules through the cell wall \cite{young_dynamics_2012}.
Here we observed negative lag times (positive $\arg{Z}$) at high amplitude and high frequency.
These values very likely do not directly translate to the biological contexts above.
Additionally, biological fluid flows tend to occur at low Reynolds numbers, meaning that inertial effects can be neglected in certain instances.
Despite these caveats, our combined experimental and theoretical study does contribute significantly to a more complete framework of characterizing dynamic fluid-structure interactions.
For biological systems subject to larger-amplitude forcing, such as neuromast cupulae responding to predator bow waves or turbulent wakes, the nonlinear regime characterized here may be more directly relevant.
Extending our study to account for lag time in the specific contexts of biological mechanosensory hairs and hair beds would shed light on the role of fluid-structure interactions in flow regulation.

In addition to biological feedback control on the short timescales of vasodilation and vasoconstriction, feedback occurs on much longer timescales through structural remodeling of the hair bed itself.
After exposing cilia to shear flow for three hours, \cite{besschetnova_identification_2010} found that cilia length decreased, while sustained mechanical stimulation also increases ciliary stiffness through a negative feedback loop in which deflection induces stiffening and shortening \cite{nguyen_primary_2015}.
Our work raises the hypothesis that ciliary remodeling maintains a value of $\tilde\omega \approx 1$.
This intermediate value corresponds to a reconfiguration regime, where changes in fluid flow velocity correspond to the greatest changes in hair height, bending moment, and thus downstream signal.
A mechanosensor operating at more extreme values of $\omega$ would lose sensitivity.
Maximum dynamic sensitivity therefore would occur near $\tilde\omega \approx 1$, suggesting that length and stiffness regulation is a mechanism by which the cell tunes its mechanosensor to operate at peak sensitivity.
This interpretation is consistent with the observed direction of regulation: higher flow rates increase $\tilde\omega$, and both the shortening \cite{besschetnova_identification_2010} and stiffening \cite{nguyen_primary_2015} of cilia under sustained flow would act to reduce $\tilde\omega$ back toward unity.
Furthermore, the drag-reducing nonlinearity means that the amplitude of stress transmitted to cell-surface mechanoreceptors saturates at high flows, providing an upper limit to mechanosensory input.

In this study, we avoided the complex problem of solving an integro-partial-differential equation through the use of the rigid rod model.
It contains one degree of freedom and constrains the hair to bend at one joint, while our prior models considered the infinite-degree-of-freedom nature of continuously bending Kirchhoff rods \cite{alvarado_nonlinear_2017, smucker_integrability_2022}. 
These higher degrees of freedom could become important at higher frequencies, but are absent for those sampled in our work.
For the intermediate Reynolds number problem, agreement between experiment and model is achieved.
Our model breaks down in the high frequency, high viscosity case, likely due to hair inertia, which is not considered in our model.
It would be interesting to compare linear models that predict measurable relaxation times — such as the Brinkman-elasticae model — with experiments in the small-amplitude regime ($\psi_0 \ll 1$).
At low Reynolds number in the elastoviscous regime, $Z'' < 0$, so the relaxation time could in principle be measured experimentally.

Besides their biological relevance, the dynamic properties of hair beds characterized here have great potential in guiding engineering applications.
One natural avenue is lab-on-chip (LOC) devices \cite{shanko_microfluidic_2019}.
Because LOC devices operate at low Reynolds number, fluid flows are laminar and mixing is diffusion-limited, posing a significant constraint.
Rigid obstacles \cite{wang_finding_2019} or grooves \cite{stroock_chaotic_2002, alam_analysis_2012-1, johnson_characterization_2002, marschewski_mixing_2015, fan_rapid_2017} placed in cross-stream flow enhance mixing by stretching and folding unmixed fluid interfaces.
Flexible hair beds are passive obstacles that produce the same folding effect \cite{wexler_bending_2013} while additionally displacing fluid vertically as hairs deflect under imposed stress.
This mechanism could further enhance mixing in the region above the hair tips.
The oscillatory response characterized here suggests that dynamic, frequency-tunable mixing may be achievable by controlling the driving frequency relative to the elastoinertial crossover.
More broadly, the physics of flexible boundaries interacting with fluid flows has inspired novel flow sensors \cite{boublil_mechanosensory_2021, brucker_feasability_2005, tao_hair_2012, glick_fluid-structure_2021}, pumps and microfluidic components \cite{zhang_metachronal_2021, zhang_metachronal_2020, masoud_harnessing_2011, alvarado_nonlinear_2017, swaminathan_bio-inspired_2013}, adhesives \cite{lee_reversible_2007, kim_biomimetic_2017}, and swimming mechanisms \cite{kwak_design_2017}.
The dynamic characterization presented here extends this foundation to time-varying flow regimes.

\section{Conclusion}

We have studied the oscillatory response of hair beds subject to low-to-intermediate Reynolds-number fluid flows.
The relationship between measured stress and imposed velocity was characterized with the mechanical impedance, whose magnitude was shown to closely follow the time-independent shear problem.
Its phase, however, was nonzero and determined by a balance between fluid inertia, elasticity, and viscous effects.
We introduced a rigid-rod model, where the hair contained a pivot anchor at the base with a nonlinear torsional spring.
We use this system as a model for naturally occurring ``hair beds" such as the endothelial glycocalyx and passive cilia. These systems sense and respond to fluid stresses, allowing them to serve as mechanosensors and flow control apparatuses.
In addition to applications in biology, hair beds pose an interesting candidate to mix fluids at low Reynolds number.

\section{Methods}
\label{sec:methods}

\paragraph{Fluid-hair system preparation.}

Hairs are manufactured via the following procedure. First, a mold is created by ablating holes through an acrylic sheet with a laser cutter. We then cast the mold with polydimethylsiloxane (PDMS) and cure in an oven overnight. For further details, please refer to \cite{nasto_air_2016}. Hair lengths are varied from $(2.6-3.77)\,\mathrm{mm}$, packing fraction from $0.042-0.216$, and fluid viscosity from $(1000-10000)\,\mathrm{cSt}$.

\paragraph{Rheology}

Rheology is done with a Discovery HR-2 rheometer. This instrument is stress-controlled, but can operate in a strain-controlled mode by employing built-in feedback mechanisms. We use the latter in order to prescribe a boundary displacement $\psi (t)=-\psi_0\cos\omega t$ which yields a strain of $\gamma(t)=k_\gamma \psi(t)$ where $k_\gamma=\frac{(R+H)^2+R^2}{(R+H)^2-R^2}=2.26$. Hairs are then coated to an inner cylinder of a concentric cylinder setup and submerged into a cup of silicone oil. Frequencies are varied from $(0.1-20)\,\mathrm{rad/s}$ and angular displacement amplitudes from $(0.1-10)\,\mathrm{rad}$. $Z_0$ and $Z_\infty\equiv Z_0\left(\frac{R}{R+L}\right)\left(\frac{(R+H)^2-(R+L)^2}{(R+H)^2-R^2}\right)$ are found by fitting the lowest frequency amplitude sweep for each hair bed with the stationary flow prediction for $\tilde{Z}$ from \cite{alvarado_nonlinear_2017}. In \cref{fig:rezs,fig:imzs}, rescaled velocities are computed with $\tilde{\omega} = \frac{4\mu L^2 v_0}{E\phi a^2 H}\left(1-\epsilon\right)^{-3/2}$.
The elastic modulus is measured using an Instron 68SC-2 tensile tester. Stress is measured as the sample is strained from $0.2$ to $0.25$ and used to compute $E$.
The experimental constants are reported in \cref{tab:vars}.
\begin{table}
\centering
    \caption{Experimental constants and their associated values \label{tab:vars} }
    \begin{tabular}{lll}
        \hline
        $H$&Channel height&$6.6\,\mathrm{mm}$\\
         $a$&Hair radius &$0.15\,\mathrm{mm}$\\
         $I$& 2nd area moment of hair's cross-section &$4.0\times 10^{-4}\,\mathrm{mm}^4$\\
        $R$& Inner cylinder radius&$15.6\,\mathrm{mm}$
    \end{tabular}
\end{table}

\paragraph{Numerics}

\Cref{eq:system} is solved using Mathematica's NDSolve function. The fluid domain is discretized into $N$ mesh segments and the resulting system of ODEs can be solved directly. An if statement toggles between the low Reynolds number flow solver (i.e. $\mathrm{Re}<.01$) and higher Reynolds number flow solver. This is necessary because the full system becomes numerically stiff as the forcing frequency goes to zero.

\section{Acknowledgements}

We thank Yuan-Nan Young, David Stein, Alp Kaya, and Jean Comtet for helpful discussions. We also thank Berkin Dortdivanlioglu for assisting in the measurement of the elastic modulus of the PDMS used to make hair beds. PJM was supported by  the US Department of Energy contract DE-FG02- 04ER54742

\bibliographystyle{apsrev}
\bibliography{hairy_rheology}

@article{alam_analysis_2012-1,
  title = {Analysis of Mixing in a Curved Microchannel with Rectangular Grooves},
  author = {Alam, Afroz and Kim, Kwang-Yong},
  year = {2012},
  month = feb,
  journal = {Chemical Engineering Journal},
  volume = {181--182},
  pages = {708--716},
  urldate = {2024-03-05}
}

@article{alvarado_nonlinear_2017,
  title = {Nonlinear Flow Response of Soft Hair Beds},
  author = {Alvarado, Jos{\'e} and Comtet, Jean and {de Langre}, Emmanuel and Hosoi, A. E.},
  year = {2017},
  month = oct,
  journal = {Nature Physics},
  volume = {13},
  number = {10},
  pages = {1014--1019},
  publisher = {{Nature Publishing Group}},
  urldate = {2021-10-29},
  copyright = {2017 Nature Publishing Group},
  langid = {english}
}

@book{audoly_elasticity_2010,
  title = {Elasticity and {{Geometry}}: {{From}} Hair Curls to the Nonlinear Response of Shells},
  shorttitle = {Elasticity and {{Geometry}}},
  author = {Audoly, Basile and Pomeau, Yves},
  year = {2010},
  month = aug,
  edition = {Illustrated edition},
  publisher = {{Oxford University Press}},
  address = {{Oxford ; New York}},
  langid = {english}
}

@article{bartosch_endothelial_2017,
  title = {Endothelial {{Glycocalyx-Mediated Nitric Oxide Production}} in {{Response}} to {{Selective AFM Pulling}}},
  author = {Bartosch, Anne Marie W. and Mathews, Rick and Tarbell, John M.},
  year = {2017},
  month = jul,
  journal = {Biophysical Journal},
  volume = {113},
  number = {1},
  pages = {101--108},
  urldate = {2022-09-14},
  langid = {english}
}

@book{bechhoefer_control_2021,
  title = {Control {{Theory}} for {{Physicists}}},
  author = {Bechhoefer, John},
  year = {2021},
  month = may,
  edition = {1st edition},
  publisher = {{Cambridge University Press}},
  address = {{New York}},
  langid = {english}
}

@article{besschetnova_identification_2010,
  title = {Identification of {{Signaling Pathways Regulating Primary Cilium Length}} and {{Flow-Mediated Adaptation}}},
  author = {Besschetnova, Tatiana Y. and {Kolpakova-Hart}, Elona and Guan, Yinghua and Zhou, Jing and Olsen, Bjorn R. and Shah, Jagesh V.},
  year = {2010},
  month = jan,
  journal = {Current Biology},
  volume = {20},
  number = {2},
  pages = {182--187},
  urldate = {2022-09-15},
  langid = {english}
}

@article{boo_flow-dependent_2003,
  title = {Flow-Dependent Regulation of Endothelial Nitric Oxide Synthase: Role of  Protein Kinases},
  shorttitle = {Flow-Dependent Regulation of Endothelial Nitric Oxide Synthase},
  author = {Boo, Yong Chool and Jo, Hanjoong},
  year = {2003},
  month = sep,
  journal = {American Journal of Physiology-Cell Physiology},
  volume = {285},
  number = {3},
  pages = {C499-C508},
  publisher = {{American Physiological Society}},
  urldate = {2022-09-15}
}

@article{boublil_mechanosensory_2021,
  title = {Mechanosensory {{Hairs}} and {{Hair-like Structures}} in the {{Animal Kingdom}}: {{Specializations}} and {{Shared Functions Serve}} to {{Inspire Technology Applications}}},
  shorttitle = {Mechanosensory {{Hairs}} and {{Hair-like Structures}} in the {{Animal Kingdom}}},
  author = {Boublil, Brittney L. and Diebold, Clarice Anna and Moss, Cynthia F.},
  year = {2021},
  month = jan,
  journal = {Sensors},
  volume = {21},
  number = {19},
  pages = {6375},
  publisher = {{Multidisciplinary Digital Publishing Institute}},
  urldate = {2022-08-31},
  copyright = {http://creativecommons.org/licenses/by/3.0/},
  langid = {english}
}

@article{brucker_feasability_2005,
  title = {Feasability Study of Wall Shear Stress Imaging Using Microstructured Surfaces with Flexible Micropillars},
  author = {Br{\"u}cker, {\relax Ch}. and Spatz, J. and Schr{\"o}der, W.},
  year = {2005},
  month = aug,
  journal = {Experiments in Fluids},
  volume = {39},
  number = {2},
  pages = {464--474},
  urldate = {2022-08-31},
  langid = {english}
}

@article{fan_rapid_2017,
  title = {Rapid Microfluidic Mixer Utilizing Sharp Corner Structures},
  author = {Fan, Liang-Liang and Zhu, Xiao-Liang and Zhao, Hong and Zhe, Jiang and Zhao, Liang},
  year = {2017},
  month = feb,
  journal = {Microfluidics and Nanofluidics},
  volume = {21},
  number = {3},
  pages = {36},
  urldate = {2022-11-30},
  langid = {english}
}

@misc{franck_understanding_2005,
  title = {Understanding {{Instrument Inertia Corrections}} in {{Oscillation}}},
  author = {Franck, A J and Instruments, {\relax TA}},
  year = {2005},
  langid = {english}
}

@article{glick_fluid-structure_2021,
  title = {Fluid-{{Structure Interaction}} of {{Flexible Whisker-Type Beams}} and {{Its Implications}} for {{Flow Sensing}} by {{Pair-Wise Correlation}}},
  author = {Glick, Raphael and Muthuramalingam, Muthukumar and Br{\"u}cker, Christoph},
  year = {2021},
  month = mar,
  journal = {Fluids},
  volume = {6},
  number = {3},
  pages = {102},
  publisher = {{Multidisciplinary Digital Publishing Institute}},
  urldate = {2022-08-31},
  copyright = {http://creativecommons.org/licenses/by/3.0/},
  langid = {english}
}

@article{gopinath_elastohydrodynamics_2011,
  title = {Elastohydrodynamics of Wet Bristles, Carpets and Brushes},
  author = {Gopinath, A. and Mahadevan, L.},
  year = {2011},
  month = jan,
  journal = {Proceedings of the Royal Society A: Mathematical, Physical and Engineering Sciences},
  volume = {467},
  number = {2130},
  pages = {1665--1685},
  publisher = {{Royal Society}},
  urldate = {2023-03-01}
}

@article{johnson_characterization_2002,
  title = {Characterization and Optimization of Slanted Well Designs for Microfluidic Mixing under Electroosmotic Flow},
  author = {Johnson, Timothy J. and Locascio, Laurie E.},
  year = {2002},
  journal = {Lab on a Chip},
  volume = {2},
  number = {3},
  pages = {135},
  urldate = {2022-07-27},
  langid = {english}
}

@article{kadekaro_centrally_2000,
  title = {Centrally {{Produced Nitric Oxide And The Regulation Of Body Fluid And Blood Pressure Homeostases}}},
  author = {Kadekaro, Massako and {Summy-Long}, Joan Y},
  year = {2000},
  journal = {Clinical and Experimental Pharmacology and Physiology},
  volume = {27},
  number = {5-6},
  pages = {450--459},
  urldate = {2022-09-13},
  langid = {english}
}

@article{kim_biomimetic_2017,
  title = {Biomimetic Wall-Shaped Adhesive Microstructure for Shear-Induced Attachment: The Effects of Pulling Angle and Preliminary Displacement},
  shorttitle = {Biomimetic Wall-Shaped Adhesive Microstructure for Shear-Induced Attachment},
  author = {Kim, Jae-Kang and Varenberg, Michael},
  year = {2017},
  month = dec,
  journal = {Journal of The Royal Society Interface},
  volume = {14},
  number = {137},
  pages = {20170832},
  publisher = {{Royal Society}},
  urldate = {2022-09-08}
}

@article{kwak_design_2017,
  title = {Design of Hair-like Appendages and Comparative Analysis on Their Coordination toward Steady and Efficient Swimming},
  author = {Kwak, Bokeon and Bae, Joonbum},
  year = {2017},
  month = may,
  journal = {Bioinspiration \& Biomimetics},
  volume = {12},
  number = {3},
  pages = {036014},
  urldate = {2022-08-30},
  langid = {english}
}

@article{landry_dynamics_2005,
  title = {Dynamics of an {{Inverted Pendulum}} with {{Delayed Feedback Control}}},
  author = {Landry, Maria and Campbell, Sue Ann and Morris, Kirsten and Aguilar, Cesar O.},
  year = {2005},
  month = jan,
  journal = {SIAM Journal on Applied Dynamical Systems},
  volume = {4},
  number = {2},
  pages = {333--351},
  publisher = {{Society for Industrial and Applied Mathematics}},
  urldate = {2024-02-07}
}

@article{lee_reversible_2007,
  title = {A Reversible Wet/Dry Adhesive Inspired by Mussels and Geckos},
  author = {Lee, Haeshin and Lee, Bruce P. and Messersmith, Phillip B.},
  year = {2007},
  month = jul,
  journal = {Nature},
  volume = {448},
  number = {7151},
  pages = {338--341},
  urldate = {2022-08-31},
  langid = {english}
}

@article{marschewski_mixing_2015,
  title = {Mixing with Herringbone-Inspired Microstructures: Overcoming the Diffusion Limit in Co-Laminar Microfluidic Devices},
  shorttitle = {Mixing with Herringbone-Inspired Microstructures},
  author = {Marschewski, Julian and Jung, Stefan and Ruch, Patrick and Prasad, Nishant and Mazzotti, Sergio and Michel, Bruno and Poulikakos, Dimos},
  year = {2015},
  journal = {Lab on a Chip},
  volume = {15},
  number = {8},
  pages = {1923--1933},
  urldate = {2022-08-16},
  langid = {english}
}

@article{martin_predator-prey_2001,
  title = {Predator-Prey Models with Delay and Prey Harvesting},
  author = {Martin, Annik and Ruan, Shigui},
  year = {2001},
  month = sep,
  journal = {Journal of Mathematical Biology},
  volume = {43},
  number = {3},
  pages = {247--267},
  urldate = {2024-02-07},
  langid = {english}
}

@article{masoud_harnessing_2011,
  title = {Harnessing Synthetic Cilia to Regulate Motion of Microparticles},
  author = {Masoud, Hassan and Alexeev, Alexander},
  year = {2011},
  journal = {Soft Matter},
  volume = {7},
  number = {19},
  pages = {8702},
  urldate = {2022-08-30},
  langid = {english}
}

@article{milton_time-delayed_2009,
  title = {The Time-Delayed Inverted Pendulum: {{Implications}} for Human Balance Control},
  shorttitle = {The Time-Delayed Inverted Pendulum},
  author = {Milton, John and Cabrera, Juan Luis and Ohira, Toru and Tajima, Shigeru and Tonosaki, Yukinori and Eurich, Christian W. and Campbell, Sue Ann},
  year = {2009},
  month = jun,
  journal = {Chaos: An Interdisciplinary Journal of Nonlinear Science},
  volume = {19},
  number = {2},
  pages = {026110},
  urldate = {2024-02-07}
}

@article{nasto_air_2016,
  title = {Air Entrainment in Hairy Surfaces},
  author = {Nasto, Alice and Regli, Marianne and Brun, P.-T. and Alvarado, Jos{\'e} and Clanet, Christophe and Hosoi, A. E.},
  year = {2016},
  month = jul,
  journal = {Physical Review Fluids},
  volume = {1},
  number = {3},
  pages = {033905},
  urldate = {2021-11-02},
  langid = {english}
}

@article{nguyen_primary_2015,
  title = {The Primary Cilium Is a Self-Adaptable, Integrating Nexus for Mechanical Stimuli and Cellular Signaling},
  author = {Nguyen, An M. and Young, Y.-N. and Jacobs, Christopher R.},
  year = {2015},
  month = dec,
  journal = {Biology Open},
  volume = {4},
  number = {12},
  pages = {1733--1738},
  urldate = {2022-10-27},
  langid = {english}
}

@article{rees_role_1989,
  title = {Role of Endothelium-Derived Nitric Oxide in the Regulation of Blood Pressure},
  author = {Rees, D. D. and Palmer, R. M. and Moncada, S.},
  year = {1989},
  month = may,
  journal = {Proceedings of the National Academy of Sciences of the United States of America},
  volume = {86},
  number = {9},
  pages = {3375--3378},
  langid = {english},
  pmcid = {PMC287135},
  pmid = {2497467}
}

@article{saternos_primary_2020,
  title = {Primary {{Cilia}} and {{Calcium Signaling Interactions}}},
  author = {Saternos, Hannah and Ley, Sidney and AbouAlaiwi, Wissam},
  year = {2020},
  month = jan,
  journal = {International Journal of Molecular Sciences},
  volume = {21},
  number = {19},
  pages = {7109},
  publisher = {{Multidisciplinary Digital Publishing Institute}},
  urldate = {2022-09-14},
  copyright = {http://creativecommons.org/licenses/by/3.0/},
  langid = {english}
}

@article{satir_overview_2007,
  title = {Overview of {{Structure}} and {{Function}} of {{Mammalian Cilia}}},
  author = {Satir, Peter and Christensen, S{\o}ren Tvorup},
  year = {2007},
  month = mar,
  journal = {Annual Review of Physiology},
  volume = {69},
  number = {1},
  pages = {377--400},
  urldate = {2022-09-14},
  langid = {english}
}

@article{shanko_microfluidic_2019,
  title = {Microfluidic {{Magnetic Mixing}} at {{Low Reynolds Numbers}} and in {{Stagnant Fluids}}},
  author = {Shanko, Eriola-Sophia and {van de Burgt}, Yoeri and Anderson, Patrick D. and {den Toonder}, Jaap M. J.},
  year = {2019},
  month = nov,
  journal = {Micromachines},
  volume = {10},
  number = {11},
  pages = {731},
  publisher = {{Multidisciplinary Digital Publishing Institute}},
  urldate = {2021-10-27},
  copyright = {http://creativecommons.org/licenses/by/3.0/},
  langid = {english}
}

@article{smucker_integrability_2022,
  title = {Integrability Technique for Fluid Flow Induced Deformation of a Boundary Hair},
  author = {Smucker, Jonas and Vural, Zerrin M. and Alvarado, Jos{\'e} R. and Morrison, Philip J.},
  year = {2022},
  month = aug,
  journal = {Physical Review Fluids},
  volume = {7},
  number = {8},
  pages = {084001},
  publisher = {{American Physical Society}},
  urldate = {2022-09-20}
}

@article{spasic_primary_2017,
  title = {Primary Cilia: {{Cell}} and Molecular Mechanosensors Directing Whole Tissue Function},
  shorttitle = {Primary Cilia},
  author = {Spasic, Milos and Jacobs, Christopher R.},
  year = {2017},
  month = nov,
  journal = {Seminars in Cell \& Developmental Biology},
  volume = {71},
  pages = {42--52},
  langid = {english},
  pmcid = {PMC5922257},
  pmid = {28843978}
}

@article{stein_coarse_2019,
  title = {Coarse Graining the Dynamics of Immersed and Driven Fiber Assemblies},
  author = {Stein, David B. and Shelley, Michael J.},
  year = {2019},
  month = jul,
  journal = {Physical Review Fluids},
  volume = {4},
  number = {7},
  pages = {073302},
  publisher = {{American Physical Society}},
  urldate = {2022-05-25}
}

@article{stroock_chaotic_2002,
  title = {Chaotic {{Mixer}} for {{Microchannels}}},
  author = {Stroock, Abraham D. and Dertinger, Stephan K. W. and Ajdari, Armand and Mezi{\'c}, Igor and Stone, Howard A. and Whitesides, George M.},
  year = {2002},
  month = jan,
  journal = {Science},
  volume = {295},
  number = {5555},
  pages = {647--651},
  publisher = {{American Association for the Advancement of Science}},
  urldate = {2022-11-30}
}

@article{swaminathan_bio-inspired_2013,
  title = {Bio-Inspired Mammalian Hair-Fabricated Microfluidics},
  author = {Swaminathan, Swathi and Harris, Thomas and McClellan, Dan and Cui, Yue},
  year = {2013},
  month = sep,
  journal = {Materials Letters},
  volume = {106},
  pages = {208--212},
  urldate = {2021-10-28},
  langid = {english}
}

@article{tao_hair_2012,
  title = {Hair Flow Sensors: From Bio-Inspiration to Bio-Mimicking{\textemdash}a Review},
  shorttitle = {Hair Flow Sensors},
  author = {Tao, Junliang and Yu, Xiong (Bill)},
  year = {2012},
  month = nov,
  journal = {Smart Materials and Structures},
  volume = {21},
  number = {11},
  pages = {113001},
  urldate = {2022-08-31},
  langid = {english}
}

@article{thomazo_collective_2020,
  title = {Collective Stiffening of Soft Hair Assemblies},
  author = {Thomazo, Jean-Baptiste and Lauga, Eric and Le R{\'e}v{\'e}rend, Benjamin and Wandersman, E. and Prevost, A. M.},
  year = {2020},
  month = jul,
  journal = {Physical Review E},
  volume = {102},
  number = {1},
  pages = {010602},
  publisher = {{American Physical Society}},
  urldate = {2023-01-19}
}

@article{ushay_interfacial_2023,
  title = {Interfacial Flows Past Arrays of Elastic Fibers},
  author = {Ushay, C. and {Jambon-Puillet}, E. and Brun, P.-T.},
  year = {2023},
  month = apr,
  journal = {Physical Review Fluids},
  volume = {8},
  number = {4},
  pages = {044001},
  urldate = {2024-02-05},
  langid = {english}
}

@article{wang_finding_2019,
  title = {Finding the Optimal Design of a Passive Microfluidic Mixer},
  author = {Wang, Junchao and Zhang, Naiyin and Chen, Jin and J.~Rodgers, Victor G. and Brisk, Philip and H.~Grover, William},
  year = {2019},
  journal = {Lab on a Chip},
  volume = {19},
  number = {21},
  pages = {3618--3627},
  publisher = {{Royal Society of Chemistry}},
  urldate = {2022-11-30},
  langid = {english}
}

@article{weinbaum_mechanotransduction_2003,
  title = {Mechanotransduction and Flow across the Endothelial Glycocalyx},
  author = {Weinbaum, Sheldon and Zhang, Xiaobing and Han, Yuefeng and Vink, Hans and Cowin, Stephen C.},
  year = {2003},
  month = jun,
  journal = {Proceedings of the National Academy of Sciences},
  volume = {100},
  number = {13},
  pages = {7988--7995},
  publisher = {{Proceedings of the National Academy of Sciences}},
  urldate = {2022-11-30}
}

@article{wexler_bending_2013,
  title = {Bending of Elastic Fibres in Viscous Flows: The Influence of Confinement{\ddag}},
  shorttitle = {Bending of Elastic Fibres in Viscous Flows},
  author = {Wexler, Jason S. and Trinh, Philippe H. and Berthet, Helene and Quennouz, Nawal and du Roure, Olivia and Huppert, Herbert E. and Lindner, Anke and Stone, Howard A.},
  year = {2013},
  month = apr,
  journal = {Journal of Fluid Mechanics},
  volume = {720},
  pages = {517--544},
  publisher = {{Cambridge University Press}},
  urldate = {2022-11-30},
  langid = {english}
}

@article{young_dynamics_2012,
  title = {Dynamics of the {{Primary Cilium}} in {{Shear Flow}}},
  author = {Young, Y. -N. and Downs, M. and Jacobs, C. R.},
  year = {2012},
  month = aug,
  journal = {Biophysical Journal},
  volume = {103},
  number = {4},
  pages = {629--639},
  urldate = {2023-03-24},
  langid = {english}
}

@article{zhang_metachronal_2020,
  title = {Metachronal Actuation of Microscopic Magnetic Artificial Cilia Generates Strong Microfluidic Pumping},
  author = {Zhang, Shuaizhong and Cui, Zhiwei and Wang, Ye and den Toonder, Jaap M. J.},
  year = {2020},
  month = sep,
  journal = {Lab on a Chip},
  volume = {20},
  number = {19},
  pages = {3569--3581},
  publisher = {{The Royal Society of Chemistry}},
  urldate = {2021-10-27},
  langid = {english}
}

@article{zhang_metachronal_2021,
  title = {Metachronal {$\mu$}-{{Cilia}} for {{On-Chip Integrated Pumps}} and {{Climbing Robots}}},
  author = {Zhang, Shuaizhong and Cui, Zhiwei and Wang, Ye and {den Toonder}, Jaap},
  year = {2021},
  month = may,
  journal = {ACS Applied Materials \& Interfaces},
  volume = {13},
  number = {17},
  pages = {20845--20857},
  publisher = {{American Chemical Society}},
  urldate = {2021-10-27}
}

@article{Boublil-2021-Sensors,
  title = {Mechanosensory {{Hairs}} and {{Hair-like Structures}} in the {{Animal Kingdom}}: {{Specializations}} and {{Shared Functions Serve}} to {{Inspire Technology Applications}}},
  shorttitle = {Mechanosensory {{Hairs}} and {{Hair-like Structures}} in the {{Animal Kingdom}}},
  author = {Boublil, Brittney L. and Diebold, Clarice Anna and Moss, Cynthia F.},
  year = 2021,
  month = jan,
  journal = {Sensors},
  volume = {21},
  number = {19},
  pages = {6375},
  publisher = {Multidisciplinary Digital Publishing Institute},
  issn = {1424-8220},
  doi = {10.3390/s21196375},
  urldate = {2026-05-12},
  copyright = {http://creativecommons.org/licenses/by/3.0/},
  langid = {english}
}

@article{Fettiplace-2014-Physiol.Rev.,
  title = {The {{Physiology}} of {{Mechanoelectrical Transduction Channels}} in {{Hearing}}},
  author = {Fettiplace, Robert and Kim, Kyunghee X.},
  date = {2014-07},
  year = 2014,
  journaltitle = {Physiological Reviews},
  volume = {94},
  number = {3},
  pages = {951--986},
  publisher = {American Physiological Society},
  issn = {0031-9333},
  doi = {10.1152/physrev.00038.2013},
  url = {https://journals.physiology.org/doi/full/10.1152/physrev.00038.2013},
  urldate = {2026-05-13}
}

@article{vanNetten-1987-HearingResearch,
  title = {Laser Interferometric Measurements on the Dynamic Behaviour of the Cupula in the Fish Lateral Line},
  author = {{van Netten}, Sietse M. and Kroese, Alfons B. A.},
  year = 1987,
  month = jan,
  journal = {Hearing Research},
  volume = {29},
  number = {1},
  pages = {55--61},
  issn = {0378-5955},
  doi = {10.1016/0378-5955(87)90205-X},
  urldate = {2026-05-13}
}

@article{McHenry-2008-JCompPhysiolA,
  title = {Mechanical Filtering by the Boundary Layer and Fluid--Structure Interaction in the Superficial Neuromast of the Fish Lateral Line System},
  author = {McHenry, Matthew J. and Strother, James A. and {van Netten}, Sietse M.},
  year = 2008,
  month = sep,
  journal = {Journal of Comparative Physiology A},
  volume = {194},
  number = {9},
  pages = {795--810},
  issn = {1432-1351},
  doi = {10.1007/s00359-008-0350-2},
  urldate = {2026-05-13},
  langid = {english}
}

@incollection{vanNetten-2014-TheLateralLineSystem,
  title = {The {{Biophysics}} of the {{Fish Lateral Line}}},
  booktitle = {The {{Lateral Line System}}},
  author = {{van Netten}, Sietse M. and McHenry, Matthew J.},
  editor = {Coombs, Sheryl and Bleckmann, Horst and Fay, Richard R. and Popper, Arthur N.},
  year = 2014,
  pages = {99--119},
  publisher = {Springer},
  address = {New York, NY},
  doi = {10.1007/2506_2013_14},
  urldate = {2026-05-13},
  isbn = {978-1-4614-8851-4},
  langid = {english}
}

@article{Nauli-2003-NatGenet,
  title = {Polycystins 1 and 2 Mediate Mechanosensation in the Primary Cilium of Kidney Cells},
  author = {Nauli, Surya M. and Alenghat, Francis J. and Luo, Ying and Williams, Eric and Vassilev, Peter and Li, Xiaogang and Elia, Andrew E. H. and Lu, Weining and Brown, Edward M. and Quinn, Stephen J. and Ingber, Donald E. and Zhou, Jing},
  year = 2003,
  month = feb,
  journal = {Nature Genetics},
  volume = {33},
  number = {2},
  pages = {129--137},
  publisher = {Nature Publishing Group},
  issn = {1546-1718},
  doi = {10.1038/ng1076},
  urldate = {2026-05-13},
  copyright = {2003 Springer Nature America, Inc.},
  langid = {english}
}

@article{Duprat-2014-LabChip,
  title = {Microfluidic in Situ Mechanical Testing of Photopolymerized Gels},
  author = {Duprat, Camille and Berthet, H{\'e}l{\`e}ne and Wexler, Jason S. and du Roure, Olivia and Lindner, Anke},
  year = 2014,
  month = dec,
  journal = {Lab on a Chip},
  volume = {15},
  number = {1},
  pages = {244--252},
  publisher = {The Royal Society of Chemistry},
  issn = {1473-0189},
  doi = {10.1039/C4LC01034E},
  urldate = {2026-05-13},
  langid = {english}
}

@article{rahimi_2022,
  title = {Drag Reduction and the {{Vogel}} Exponent of a Flexible Beam in Transient Shear Flows},
  author = {Rahimi, Ali Mehdizadeh and Lustig, Steven R. and Bardhan, Jaydeep P. and Jamali, Safa},
  year = 2022,
  month = oct,
  journal = {Physics of Fluids},
  volume = {34},
  number = {10},
  pages = {104111},
  issn = {1070-6631},
  doi = {10.1063/5.0106700},
  urldate = {2026-01-29}
}

@article{Pang-2025-PhysicsofFluids,
  title = {Nonlinear Flow Response of Flexible Fibers Driven by {{Stokes}} Flow},
  author = {Pang, Bo (庞博) and Sun, Bo Hua (孙博华)},
  year = 2025,
  month = feb,
  journal = {Physics of Fluids},
  volume = {37},
  number = {2},
  pages = {022133},
  issn = {1070-6631},
  doi = {10.1063/5.0251611},
  urldate = {2026-01-29}
}

@article{Sun-2025-PhysicsofFluids,
  title = {Numerical Algorithm for Nonlinear Flow Response of Soft Brush Beds Based on the Homotopy Method},
  author = {Sun, Bo Hua (孙博华) and Li, Meng (李蒙) and Pang, Bo (庞博)},
  year = 2025,
  month = jan,
  journal = {Physics of Fluids},
  volume = {37},
  number = {1},
  pages = {012122},
  issn = {1070-6631},
  doi = {10.1063/5.0249844},
  urldate = {2026-01-29}
}

@article{Peng-2021-ExpMecha,
  title = {Bending of {{Soft Micropatterns}} in {{Elastohydrodynamic Lubrication Tribology}}},
  author = {Peng, Y. and Serfass, C. M. and Hill, C. N. and Hsiao, L. C.},
  year = 2021,
  month = jul,
  journal = {Experimental Mechanics},
  volume = {61},
  number = {6},
  pages = {969--979},
  issn = {1741-2765},
  doi = {10.1007/s11340-021-00715-8},
  urldate = {2026-01-29},
  langid = {english}
}

@article{Jambon-Puillet-2026-J.FluidMech.,
  title = {Hydraulic Resistance of Channels Obstructed by a Dense Array of Elastic Fibres},
  author = {{Jambon-Puillet}, Etienne},
  year = 2026,
  month = feb,
  journal = {Journal of Fluid Mechanics},
  volume = {1028},
  pages = {A21},
  issn = {0022-1120, 1469-7645},
  doi = {10.1017/jfm.2025.11060},
  urldate = {2026-05-06},
  langid = {english}
}

@misc{Jammalamadaka-2026-,
  title = {Nonlinear Response of Soft Hair Beds to {{Poiseuille}} Flows},
  author = {Jammalamadaka, Mani Sai Suryateja and Smucker, Jonas and Alvarado, Jose R.},
  year = 2026,
  month = apr,
  number = {arXiv:2604.03804},
  eprint = {2604.03804},
  primaryclass = {physics},
  publisher = {arXiv},
  doi = {10.48550/arXiv.2604.03804},
  urldate = {2026-05-06},
  archiveprefix = {arXiv}
}

@article{mitsoulas_2022,
  title = {Dynamics and Apparent Permeability of the Glycocalyx Layer: {{Start-up}} and Pulsating Shear Experiments in Silico},
  shorttitle = {Dynamics and Apparent Permeability of the Glycocalyx Layer},
  author = {Mitsoulas, Vlasis and Varchanis, Stylianos and Dimakopoulos, Yannis and Tsamopoulos, John},
  year = 2022,
  month = jan,
  journal = {Physical Review Fluids},
  volume = {7},
  number = {1},
  pages = {013102},
  publisher = {American Physical Society},
  doi = {10.1103/PhysRevFluids.7.013102},
  urldate = {2026-01-29}
}

@article{McHenry-2007-JExpBiol,
  title = {The Flexural Stiffness of Superficial Neuromasts in the Zebrafish({{Danio}} Rerio) Lateral Line},
  author = {McHenry, Matthew J. and {van Netten}, Sietse M.},
  year = 2007,
  month = dec,
  journal = {Journal of Experimental Biology},
  volume = {210},
  number = {23},
  pages = {4244--4253},
  issn = {0022-0949},
  doi = {10.1242/jeb.009290},
  urldate = {2026-05-13}
}

@article{thomazoProbingInmouthTexture2019,
  title = {Probing In-Mouth Texture Perception with a Biomimetic Tongue},
  author = {Thomazo, Jean-Baptiste and Pastenes, Javier Contreras and Pipe, Christopher J. and R{\'e}v{\'e}rend, Benjamin Le and Wandersman, Elie and Prevost, Alexis M.},
  year = 2019,
  journal = {Journal of the Royal Society Interface},
  volume = {16},
  number = {159},
  pages = {20190362},
  issn = {1742-5689},
  doi = {10.1098/rsif.2019.0362}
}

\FloatBarrier
\section{Supplementary Material}
\beginsupplement

\begin{figure}
    \centering
    \includegraphics[scale=.7]{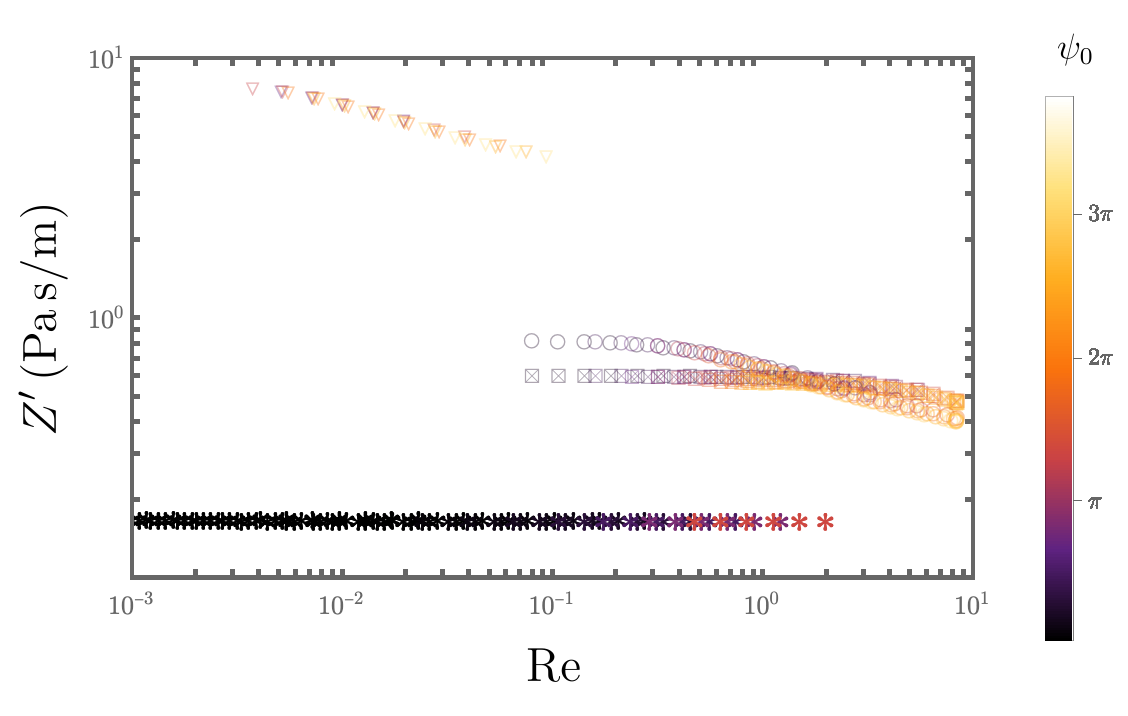}
   \caption{Plot of the real part $Z'$ of the impedance as a function of Reynolds number $\mathrm{Re}\equiv\frac{\rho \omega \psi_0 R}{\mu}$. Stars: control experiment with a rotor lacking hairs. All other symbols: experiments, as presented in the main text (cf. \cref{fig:rezs}). For clarity, the opacity of the hair-bed experiments has been decreased.}
   \label{fig:reZControl}
\end{figure}

\begin{figure}
    \centering
    \includegraphics[scale=.7]{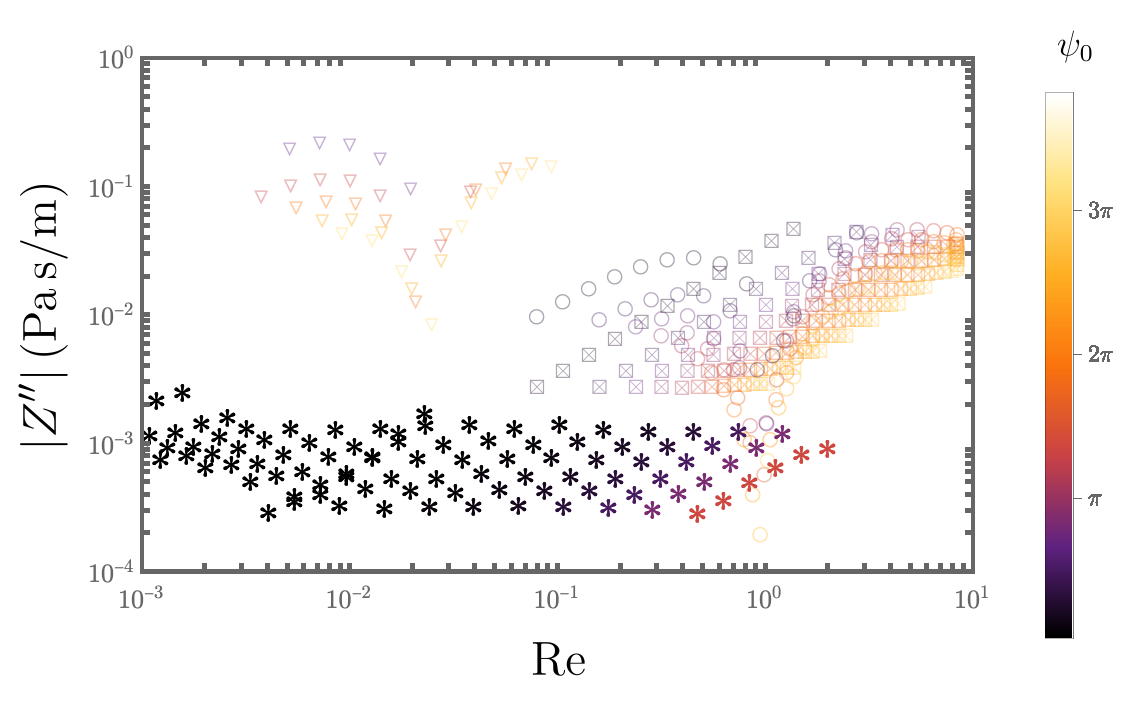}
   \caption{Plot of the imaginary part $Z''$ of the impedance as a function of Reynolds number $\mathrm{Re}\equiv\frac{\rho \omega \psi_0 R}{\mu}$. Stars: control experiment with a rotor lacking hairs. All other symbols: experiments, as presented in the main text (cf. \cref{fig:rezs}). For clarity, the opacity of the hair-bed experiments has been decreased.}
   \label{fig:imZControl}
\end{figure}

\begin{figure}
    \centering
    \includegraphics[scale=.7]{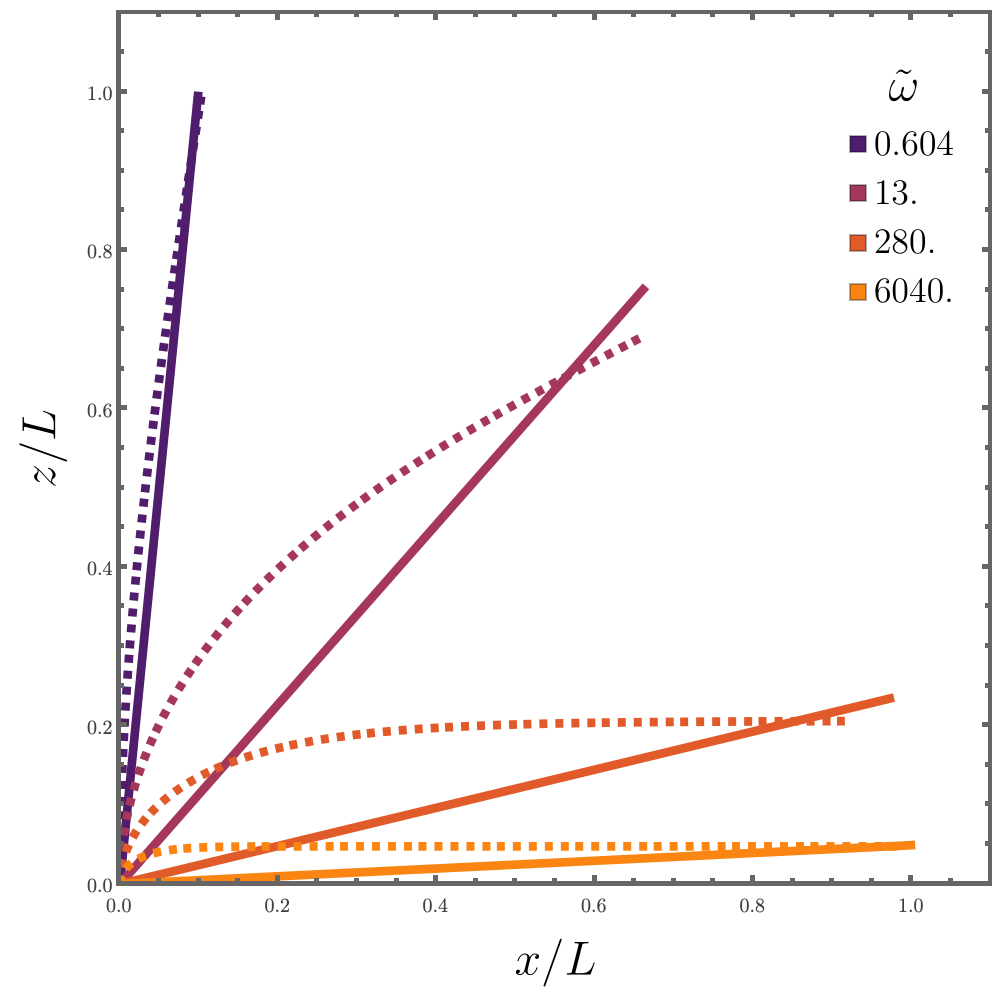}
   \caption{Comparison of profiles between flexible hairs (dashed lines) and rigid rods (solid lines) for different values of the rescaled frequency $\tilde \omega$ (legend). }
   \label{fig:prfls}
\end{figure}

\subsection{Inertial Effects}
\Cref{fig:reZControl,fig:imZControl} show the control in comparison with the hair-bed experiments. 
In probing high frequencies, one artifact that may occur results from the inertia of the rotor of stress-controlled rheometers. This inertia can be subtracted off to a certain extent, but will cause artifacts at high frequency due to the finite precision of our measurement. This effect can be quantified by the ratio of inertia to sample torque, $\mathcal{I}$ \cite{franck_understanding_2005}:
\begin{equation}
\mathcal{I}\equiv\frac{\delta I k_\tau\omega}{\mu},
\end{equation}
where $\delta I\sim \qty{E-2}{\micro\newton\meter\second^2}$ is the measurement uncertainty in the rotor's inertia. This ratio, $\mathcal{I}$ is of order $10^{-3}$ at most in our experiments. Hence, the effects of instrument inertia are negligible after performing the correction discussed above.

\subsection{Nonlinear spring constant}

In this section, we wish to express nonlinear coupling between shear flows and hair deformation in the context of the simplified model presented in the  main text. In this model, we assume a rigid rod pinned at the base with a nonlinear torsional spring. The restoring torque $k(\hat v) \sin \theta$ depends on the dimensionless forcing parameter $\hat v = \frac{4 \mu L^2 v}{E H \phi a^2}$. The parameter $\mu$ is the fluid viscosity, $L$ the hair length, $v$ the velocity of the opposing surface driving fluid flow, $E$ the Young’s modulus of the hair, $H$ the channel height, $\phi$ the hair bed packing fraction, and $a$ the hair radius. In order to determine the form of the nonlinear spring constant $k(\hat v)$ that approximates nonlinear coupling, we postulate a form $k(\hat v) = k_0(\hat v) + k_1(\hat v)$. In the low forcing limit $\hat v \rightarrow 0$ we expect $k \rightarrow k_0(\hat v)$, while $k \rightarrow k_1(\hat v)$ dominates at strong forcing $\hat v \rightarrow \infty$.

To determine $k_0$ and $k_1$, we calculate the height $h = \int_0^1 d\sigma \, \cos(\theta(\sigma))$ of the tip of the flexible hair (“F”) and the rigid rod (“R”) in the weak forcing (“0”) and strong forcing (“1”) limits. Here, $\sigma = s/L$ is the non-dimensionalized curvilinear coordinate. The angle of flexible hairs $\theta(\sigma)$ is given by our previous work. For the rigid rod model, we recover $\theta$ by algebraically solving torque balance equations to second order. By setting heights equal for the weak and strong forcing cases separately ($h_{\rm F0} = h_{\rm R0}$ and $h_{\rm F1} = h_{\rm R1}$), we recover expressions for $k_0$ and $k_1$ in terms of the forcing parameter $\hat v$.

In addition, we non-dimensionalize spring constants by the natural scale $k^* = Ea^4L^{-1}$ resulting from the restoring bending moment of flexible hairs.

\paragraph{Flexible hair in the weak forcing limit.} We have solved this problem in our previous study (See \cite{alvarado_nonlinear_2017}, Supplementary Information, “Perturbation analysis of the equation of equilibrium”). In short, we solved the equation
$$0 =\theta''(\sigma) \left( 1- \epsilon \int_0^1 \cos(\theta(\sigma^\backprime) \, d\sigma^\backprime \right) + \hat v \cos \theta(\sigma).$$
The parameter $\epsilon = L/H$ relates hair length $L$ to the channel height $H$. Solving for $\theta = \sum_n \hat v^{n} \theta_{(n)}(\sigma)$ to second order yielded
$$\frac{h_{\rm F0}}{L} = 1 - \frac{1}{15} \left( \frac{\hat v}{1-\epsilon} \right)^2.$$

\paragraph{Rigid rod in the weak forcing limit.} At low forcing, torque balance gives
$$k_0 \sin \theta = fL \cos \theta,$$
with $f = \frac{\pi a^2 \mu v}{H \phi (1-\epsilon)}$ the applied force due to fluid shear stresses. Expanding to second order in the small parameters $\theta$ and $fL/k_0$ yields the solution $\theta = fL/k_0$. Recognizing $\frac{\hat v}{1-\epsilon} = \frac{\pi }{4} \frac{fL}{k^*}$, we recover the rod height as
$$\frac{h_{\rm R0}}{L} = \cos \theta = 1 - \frac{\pi^2}{2^5} \left( \frac{\hat v}{1-\epsilon}\frac{1}{k_0/k^*} \right)^2.$$

\paragraph{Flexible hair in the strong forcing limit.} In the opposite limit of high forcing, we turn to our analytical solution \cite{smucker_integrability_2022}. In this limit, hairs adopt an approximately circular shape at the base and remain straight with $\theta = \pi/2$. The solution for $\theta(\sigma)$ therefore reduces in this limit to
$$\theta(\sigma) = \begin{cases} \theta'_0 \sigma & 0 \leq \sigma \leq \sigma^* \\ \pi/2 & \sigma^* \leq \sigma \leq 1. \end{cases}$$
Determining the effective hair height $h=\sigma^* L$ requires calculating the curvature $\theta’_0$ at the base. Using \cite{smucker_integrability_2022}, (eq. 22, with $\omega_\epsilon^2 \rightarrow \omega^2 \equiv \hat v, \xi \rightarrow 1, k \rightarrow 1$)
$$\theta'_0 = \theta'(\sigma=0) = 2 \hat v^{1/2} \, \mathbf{dn}(\mathbf{F}(\tfrac{\pi}{4})) = \sqrt{2} \hat v^{1/2},$$
where $\mathbf{dn}$ denotes the Jacobi delta amplitude and $\mathbf{F}$ the incomplete elliptic integral of the first kind. With $\theta(\sigma^*) = \sigma^*\theta'_0 = \tfrac{\pi}{2}$, we recover

$$\frac{h_{\rm F1}}{L} = \frac{\pi}{2^{3/2}}\frac{1}{\hat v^{1/2}}.$$

\paragraph{Rigid rod in the strong forcing limit.} In this limit, the driving force becomes $f = \frac{\pi a^2 \mu v}{H \phi}$. We solve the torque balance equation, but this time to second order in the small parameters $\delta = \frac{\pi}{2}-\theta$ and $\left(fL/k_1\right)^{-1}$. The solution, $\delta = \left(fL/k_1\right)^{-1}$, yields

$$\frac{h_{\rm R1}}{L} = \sin{\delta} = \frac{k_1(\hat v)}{k^*} \frac{1}{\hat v}.$$

\paragraph{Nonlinear spring constant.} Equating $h_{\rm F0} = h_{\rm R0}$ yields the low-forcing spring constant

$$\frac{k_0}{k^*} = \frac{\sqrt{15}}{2^{5/2}},$$
which is independent of $v$ to second order. Meanwhile, we recover the high-forcing spring constant by equating $h_{\rm F1} = h_{\rm R1}$, yielding

$$\frac{k_1}{k^*} = \frac{\pi}{2^{3/2}} \hat v^{1/2}.$$
We therefore find a $v^{1/2}$-dependence of the spring constant at high forcing. Putting both terms together, we recover

$$\frac{k(\hat v)}{Ea^4L^{-1}} = \frac{\pi}{2^{3/2}} \left(\frac{\sqrt{15}}{2} +  \hat v^{1/2} \right).$$
    
\end{document}